\documentclass[useAMS,usenatbib,referee]{mn2e}

\usepackage{graphicx}
\usepackage{epsfig}
\usepackage{subfigure}
\usepackage{amsmath}
\usepackage{amssymb}

\def\kms{\,{\rm km~s^{-1}}}
\def\mso{\,{\rm M}_\odot}

\def\msoy{\,{{\rm M}_\odot~{\rm yr}^{-1}}}

\title[Supernova collisions with circumstellar shells]{Numerical
  models of collisions between core-collapse supernovae and
  circumstellar shells} 
\author[A. J. van Marle et al.] {Allard Jan van Marle$^{1,2}$, Nathan
  Smith$^3$, Stanley P.\ Owocki$^2$ and B. van Veelen $^4$\\ 
  $^1$ Centre for Plasma Astrophysics, K.U. Leuven, 
  Celestijnenlaan 200B, B-3001, Leuven, Belgium\\ 
  $^2$Bartol Research Institute, University of Delaware, Newark, DE
  19716, USA \\ $^2$Astronomy Department, University of California,
  601 Campbell Hall, Berkeley, CA 94720 \\ 
  $^4$ Astronomical Institute, Utrecht University, P.O. Box
  80\,000, 3508 TA Utrecht, the Netherlands
  } \date{Submitted ??/??}
\pagerange{\pageref{firstpage}--\pageref{lastpage}} \pubyear{2009}

\begin{document}

\maketitle

\label{firstpage}

\begin{abstract}

  Recent observations of luminous Type IIn supernovae (SNe) provide
  compelling evidence that massive circumstellar shells surround their
  progenitors.  In this paper we investigate how the properties of
  such shells influence the SN lightcurve by conducting numerical
  simulations of the interaction between an expanding SN and a
  circumstellar shell ejected a few years prior to core collapse.  Our
  parameter study explores how the emergent luminosity depends on a
  range of circumstellar shell masses, velocities, geometries, and
  wind mass-loss rates, as well as variations in the SN mass and
  energy.  We find that the shell mass is the most important
  parameter, in the sense that higher shell masses (or higher ratios
  of M$_{shell}$/M$_{SN}$) lead to higher peak luminosities and higher
  efficiencies in converting shock energy into visual light.  Lower
  mass shells can also cause high peak luminosities if the shell is
  slow or if the SN ejecta are very fast, but only for a short time.
  Sustaining a high luminosity for durations of more than 100 d
  requires massive circumstellar shells of order 10~$M_{\odot}$ or
  more.  This reaffirms previous comparisons between pre-SN shells and
  shells produced by giant eruptions of luminous blue variables
  (LBVs), although the physical mechanism responsible for these
  outbursts remains uncertain.  The lightcurve shape and observed
  shell velocity can help diagnose the approximate size and density of
  the circumstellar shell, and it may be possible to distinguish
  between spherical and bipolar shells with multiwavelength
  lightcurves.  These models are merely illustrative.  One can, of
  course, achieve even higher luminosities and longer duration light
  curves from interaction by increasing the explosion energy and shell
  mass beyond values adopted here.

\end{abstract}

\keywords{hydrodynamics --- methods: numerical --- stars: mass loss
  --- stars: supernovae (general) --- stars: winds, outflows }

\section{Introduction}

The luminosity of a supernova (SN) results from energy input by a
combination of radioactive decay and shock kinetic energy 
\citep[see e.g.,][]{A96}, and for a Type II SN, the shape of the light curve
depends on quantities like the star's initial radius, ejecta mass, and
explosion energy \citep{A96, Y04, KW09}.  For SNe with small initial
radii, like SNe of Types Ia, Ib, Ic, and peculiar SNe~II like SN~1987A
that result from blue supergiants, most of the shock-deposited thermal
energy imparted to the stellar envelope is converted back into kinetic
energy through adiabatic expansion, so nearly all of the observed
luminosity comes from the radioactive decay of $^{56}$Ni and
$^{56}$Co.  In ``normal'' SNe~II-P that result from the explosions of
red supergiants (RSGs), however, the large initial radius allows some
modest fraction (typically 1--2\%) of the shock-deposited thermal
energy to be radiated away, powering much of the plateau of the
lightcurve, although the vast majority still goes into expansion
energy.  At late times, even SNe~II-P have their luminosity powered by
radioactive decay (e.g., \citealt{H03}).

Subsequently, however, as the fast SN ejecta expand, they can collide
with dense circumstellar or interstellar material (CSM/ISM) that may
surround the SN.  As a result, additional kinetic energy may be
transformed once again back into thermal energy through shock heating,
which in turn may be lost by radiative cooling if a dense radiative
shock forms (e.g., \citealt{CF08}).  This can enhance the luminosity
for long after the explosion: Som SNe remain radio luminous for
decades \citep{M98, W02, V93}, and this interaction may power a
visible supernova remnant (SNR) such as Cas~A for hundreds of years
\citep{C77, CO03}.  On the other hand, if the collision with dense CSM
happens immediately after the explosion, it may significantly alter
the spectrum and light curve of the SN itself.  This latter scenario
is generally thought to be the case for the observed sub-class of
Type~IIn supernovae \citep{S90,F97}, where the ``n'' corresponds to
``narrow'' or intermediate-width H lines from the shock-heated CSM gas
or decelerated SN ejecta (e.g. \citealt{CD94,C01}).

In a normal SN, the expected results of radiative cooling and
reheating of the SN ejecta due to radioactive decay yield can be
estimated from analytical models of stellar structure and explosion
physics \citep{MM99}.  In SNe with strong CSM interaction such as the
observed class of Type~IIn SNe, however, the effects of collisions
between an expanding SN and its circumstellar gas are harder to
predict with {\it ab initio} calculations.  They depend highly on the
density and morphology of the CSM, which in turn depend on the unknown
mass-loss behavior of the star in the few years prior to core collapse
--- potentially different for each object.  A wide variety of CSM
environments are possible, leading to a wide diversity of observed
lightcurves and spectral properties.

Recent observations of luminous Type~IIn supernovae such as SN~2006gy
\citep{S07,O07} and SN~2006tf \citep{S08a} have stretched the
boundaries of our understanding of SNe~IIn.  Their extreme
luminosities yield strong evidence that the progenitors of these SNe
were surrounded by massive shells, presumably ejected in precursor
eruptions during the final years of stellar evolution \citep{S07,
  S08a, S09b, SM07, WBH07}.  \citet{S07} pointed out that the physical
properties (mass, speed, H composition) of these mass ejections were
analogous to those observed for giant eruptions of luminous blue
variables (LBVs), and especially reminiscent of the giant 1843
eruption of $\eta$ Carinae \citep{S03}.  As the SN ejecta expand, they
collide with the recently ejected CSM shell and this collision
significantly decelerates the SN expansion, transforming kinetic
energy back into thermal energy at the collision front, producing a
brilliant fireworks display.  The remarkably high luminosity and long
duration of the observed emission from SNe~2006gy and 2006tf imply
that the circumstellar shells were very massive --- of order 10--20
$M_{\odot}$ --- in order to sufficiently decelerate the SN blast wave
and tap into its available reservoir of kinetic energy
\citep{S07,S08a,S09b,SM07,WBH07}.

\citet{SM07} have argued based on a simplified analytical model,
similar to that of \citet{FA77}, that the high luminosity and long
duration of these SNe can be explained by a SN colliding with a very
massive and initially opaque CSM shell.  We explore this idea here in
more detail with a variety of possible CSM environments using
numerical simulations.  We suggest that the presence and shape of
circumstellar shells can be a powerful tool to constrain the evolution
of the progenitors of Type~IIn supernovae.  We investigate how the
mass, speed, and morphology of such shells can influence the evolution
of a SN lightcurve.  We undertake a parameter study of SNe with
different masses interacting with a selection of possible
circumstellar shells, both spherical and bipolar.  From these
simulations we calculate thermal emission profiles and compare them in
order to constrain how the physical properties of circumstellar
nebulae can influence the SN lightcurve, and to constrain the
efficiency of converting kinetic energy to light.

Our calculations are simplified in the way we treat the cooling of and
radiation from the shocked gas, which we approximate as optically thin
radiative cooling; by necessity; our application of these results is
therefore limited in scope.  An important point to note is that our
approach is to simulate a variety of hypothetical SNe to demonstrate
{\it trends} in how the lightcurve responds to changes in SN and shell
properties.  We are not attempting a quantitative fit to the observed
data for any individual SN.  This has been pursued for a few
relatively nearby and well-observed SNe IIn, such as SN~1988Z
\citep{T93,CD94,Aetal99}, SN~1994W \citep{Cetal04}, and SN~1998S
\citep{C01}, where the CSM properties were derived from fitting the
observed light curves and spectra.  Those authors inferred massive
precursor shell ejections in the few years before core collapse,
although the energy demands and required shell masses for these were
not as extreme as for SNe~2006tf and 2006gy.  Our work here builds
upon these earlier studies.

\smallskip

We explain our adopted initial conditions and the numerical method in
\S 2 and \S 3, respectively, and in \S 4 we discuss some details of
the shock interaction.  In \S 5 we discuss how the resulting light
curves depend on various parameters and in \S 6 we discuss shell
velocities, and how these may help to interpret observations.
Finally, in \S 7 we interpret our results in context with the most
luminous SNe IIn, and in \S8 we provide a summary.

We include electronic datafiles containing the results of our
simulations with this paper.  The L\_....dat files contain the total
luminosity [erg/s] as a function of time [s].  The V\_....dat files
contain both the volume averaged and mass averaged velocity of the
shocked gas [cm/s] as a function of time [s].  A small sample of these
tables is provided in Appendices \ref{sec-lum_table} and
\ref{sec-vel_table}.

\begin{table*}
   \label{tab:sim}
    \begin{centering}
      \caption{Simulation input parameters}
      \begin{tabular}{p{0.1\linewidth}ccccccccl}
         \hline
         \noalign{\smallskip} Name
 & $M_{\rm sn}$   & $E_{\rm sn}$  & $M_{\rm shell}$  & $V(\theta=0)$ & $\dot{M}_{\rm wind}$ & $\Omega$ & t$_{\rm end}$ & $dE/E_{sn}$ & $v_{final}^{1}$ [km/s] \\
 & [$\mso$]     &  $10^{51}$~ergs&  [$\mso$]        & [$\kms$]      & [$\msoy$]            & $\Omega$ &  [yr pre-SN]  &  \% &  $10^3$ [km/s]\\
         \noalign{\smallskip}
         \hline
         \noalign{\smallskip}
                O01  & 30 & 1 & N/A &200 &  $10^{-4}$ &  0.0 & N/A & 0.05 & 4.39 \\
                O02  & 30 & 1 & N/A &200 &  $10^{-3}$ &  0.0 & N/A & 0.3  & 3.42 \\		
                O03  & 30 & 1 & N/A &200 &  $10^{-2}$ &  0.0 & N/A & 1.5  & 2.57 \\		
                O04  & 30 & 1 & N/A & 50 &  $10^{-4}$ &  0.0 & N/A & 0.108 & 3.75 \\		
	 \hline	                
	A00  & 30 & 1  &0.1 & 200 &  $10^{-4}$ &  0.0 & 2  & 0.8  & 2.85 \\
                A01  & 30 & 1  & 1  & 200 &  $10^{-4}$ &  0.0 & 2  & 5.05 & 2.34 \\
                A02  & 30 & 1  & 6  & 200 &  $10^{-4}$ &  0.0 & 2  & 18.7 & 1.79 \\
                A03  & 30 & 1  & 10 & 200 &  $10^{-4}$ &  0.0 & 2  & 25.5 & 1.59 \\
                A04  & 30 & 1  & 20 & 200 &  $10^{-4}$ &  0.0 & 2  & 36.5 & 1.27 \\
                A05  & 30 & 1  & 10 & 200 &  $10^{-3}$ &  0.0 & 2  & 25.6 & 1.51 \\
                A06  & 30 & 1  & 10 & 200 &  $10^{-5}$ &  0.0 & 2  & 25.3 & 1.61 \\
                A07  & 30 & 1  & 10 &  50 &  $10^{-3}$ &  0.0 & 2  & 31.5 & 1.30 \\
                A08  & 30 & 1  & 10 & 500 &  $10^{-3}$ &  0.0 & 2  & 16.9 & 1.66 (at500 days)\\
                A09  & 30 & 1  & 10 &  50 &  $10^{-4}$ &  0.0 & 2  & 31.6 & 1.36 \\
                A10  & 30 & 1  & 10 & 500 &  $10^{-4}$ &  0.0 & 2  & 16.3 & 1.80 (at500 days)\\
                A11  & 30 & 1  & 10 &  50 &  $10^{-5}$ &  0.0 & 2  & 31.6 & 1.46 \\
                A12  & 30 & 1  & 10 & 500 &  $10^{-5}$ &  0.0 & 2  & 16.3 & 1.80 (at 500 days)\\
          \hline
                B01  & 10 & 1  & 10 &  50 &  $10^{-4}$ &  0.0 & 2 & 54.5  & 1.52 \\
                B02  & 10 & 1  & 10 & 200 &  $10^{-4}$ &  0.0 & 2 & 48.2  & 1.83 \\
                B03  & 10 & 1  & 10 & 500 &  $10^{-4}$ &  0.0 & 2 & 37.0  & 2.12 (at 500 days) \\
                B04  & 10 & 1  & 25 & 200 &  $10^{-4}$ &  0.0 & 2 ($\Delta t=5$yr) & 65.1 &  1.08 (at 500 days) \\ % with larger outer radius, and delta t = 5 yr instead of 2
          \hline
                C01  & 60 & 1  & 10 &  50 &  $10^{-4}$ &  0.0 & 2 & 19.7 & 1.18 \\
                C02  & 60 & 1  & 10 & 200 &  $10^{-4}$ &  0.0 & 2 & 14.5 & 1.30 \\
                C03  & 60 & 1  & 10 & 500 &  $10^{-4}$ &  0.0 & 2 & 7.55 & 1.46 \\
          \hline
                D01  & 10 & 1  & 10 & 500 &  $10^{-4}$ &  0.9 & 2 & 42.1 &  N/A \\
                D02  & 30 & 1  & 10 & 500 &  $10^{-4}$ &  0.9 & 2 & 20.4 &  N/A \\
                D03  & 60 & 1  & 10 & 500 &  $10^{-4}$ &  0.9 & 2 & 10.9 &  N/A \\
          \hline
                E01  & 30 & 1  & 10 & 500 &  $10^{-4}$ &  0.0 & 4  & 14.7 & 1.79 (at 500 days) \\
                E02  & 30 & 1  & 10 & 200 &  $10^{-4}$ &  0.0 & 10 & 24.3 & 1.62 (at 500 days) \\
                E03  & 30 & 1  & 10 & 500 &  $10^{-4}$ &  0.0 & 10 & 13.9 & 1.79 (at 1000 days) \\
          \hline
                F01  & 30 & 0.5 & 10 & 200 &  $10^{-4}$ &  0.0 & 2 & 22.3 & 1.13 \\
                F02  & 30 & 2   & 10 & 200 &  $10^{-4}$ &  0.0 & 2 & 27.4 & 2.19 \\
          \hline
                G01  &  6 & 1   &  6 & 200 &  $10^{-5}$ &  0.0 & 2 & 56.4 & 2.07 (at 100 days)\\
          \hline
                H01  &  1 & 1   &  1 & 200 &  $10^{-5}$ &  0.0 & 2 & 42.7 & 4.77 (at 100 days)\\ 
          \hline
         \noalign{\smallskip}
     \end{tabular}
  \end{centering}
      $(1)$ Measured at 250 days unless indicated otherwise. \\
\end{table*}

\section{Initial conditions}

\subsection{Supernova model}\label{sec-snmodel}

In our simulations, we begin with a core-collapse SN in free expansion
as described by \citet{CF92}, \citet{MM99} and \citet{C05}, which
gives a density profile divided in two segments: The inner part has
$\rho \sim r^{-m}$, the outer part $\rho \sim r^{-b}$, with $m=1.06$
and $b=11.7$ for a progenitor star that still has a large hydrogen
envelope at the moment of core collapse.  The division between the two
power laws lies at the transition velocity:

\begin{equation}
\begin{aligned}
v_{\rm tr}~&=~3160 \sqrt{\biggl(\frac{(5-m)(b-5)}{(3-m)(b-3)}\biggr)} \\
           &\times~ \sqrt{E_{51} \biggl(\frac{10\mso}{M_{\rm ej}}\biggr)}\,[\kms],
\end{aligned}
\end{equation}

\noindent \citep{CF92,C05}. Using this profile we construct three SNe,
with different mass but equal energy.  Because of the large value of
$b$, the density drops very quickly at higher velocities.  As a
result, only a small fraction of the mass is moving fast, limiting the
inertia.  Therefore, the gas will slow down quickly when it collides
with the circumstellar medium (CSM).  Our standard massive-star SN has
30 $\mso$ of ejecta mass and 10$^{51}$ erg of total kinetic energy,
although we explored a range of SN ejection masses at 6, 10, 30 and
60~$\mso$, with total kinetic energies of 0.5, 1, and
2$\times$10$^{51}$~erg.  We start each simulation of the CSM
interaction at the moment were the supernova has expanded to 1~AU.
Typical maximum velocity for the initial supernova is about
30\,000~km/s. However, at this velocity the density is very low and
the maximum velocity is quickly reduced to about 10\,000~km/s by the
collision with the surrounding medium, once the simulation begins.
The distance it must travel to collide with the shell depends on the
shell parameters (see \S~\ref{sec-csmmodel}).

Our simulations do not include the effect of photo-ionization, nor do
we take into account the effect of energy injection from radioactive
decay.  Our calculations simulate only the expected luminosity
generated by the SN-CSM interaction shock front; our simulated light
curves do not include emission from the expanding SN photosphere
powered by diffusion of shock-deposited energy or from radioactive
decay.  These may affect the light curve at lower CSM-interaction
luminosities or very early times before the shock overtakes much of
the CSM shell.  Note also, that our supernova model is strictly
spherical.  Non-spherical SN ejecta outflows would greatly increase
the parameter space and require a more complex calculation.

\subsection{Circumstellar shell model}\label{sec-csmmodel}

For the circumstellar shells, we take a variety of shell properties,
but we focus on models reminiscent of the environment of $\eta$
Carinae \citep{S06,S03}, as such CSM properties have been proposed for
some luminous SN~IIn.  Namely, we adopt a stellar wind with high
mass-loss rate (10$^{-5}$ to 10$^{-3}$ $M_{\odot}$ yr$^{-1}$) and
moderate velocity (few hundred km s$^{-1}$) for the steady wind phase
before and after shell ejection, plus an expanding shell with
extremely high density that was ejected in an intermittent outburst
reminiscent of giant LBV eruptions, occuring a few years before the
SN.  The mass-loss rate and velocity of the wind before and after the
shell ejection are assumed to be identical.  The supernova will
therefore first encounter a (relatively) low density wind, then a
short stretch of high density material and then once again the low
density wind after it escapes the shell.  We explore a large parameter
space, covering a wide range of possible shell masses, velocities,
wind mass-loss rates and ages.

We also investigate the effect of a bipolar shell, such as might be
ejected by a rapidly rotating star (e.g.,
\citealt{DO02,o05,ST07}). The bipolar shape follows the gravitational
darkening model for the wind of a rotating star as described by
\citet{DO02}:

\begin{equation}
\frac{\dot{M}(\theta)}{\dot{M}(0)}~=~1-\Omega^2 \sin^2(\theta),
\label{eq:mdot}
\end{equation}
\begin{equation}
\frac{v_\infty(\theta)}{v_\infty(0)}~=~\sqrt{1-\Omega^2 \sin^2(\theta)},
\label{eq:vinf}
\end{equation}

\noindent with $\Omega\equiv\omega/\omega_{\rm c}$,
$\omega=\sqrt{g/R}$ the rotational angular velocity of the star and
$\omega_{\rm c}$ the Kepplerian angluar velocity.  Observations have
shown that the bipolar shell of $\eta$~Carinae, for example, follows
this shape \citep{S06}. The latitudinal angle $\theta$ equals zero at
the pole and $90^o$ at the equator.  Note that this set of equations
only applies for radiatively driven winds.  Should the star approach
critical rotation during an eruption, mass could be focussed to the
equator, forming a flattened equatorial structure as well
\citep{ST07}.  This is not accounted for in these equations.

The total range of parameters in our simulations is listed in
Table~\ref{tab:sim}.  In all cases we assume that the shell ejection
lasted two years, though we explore the effect of different shell
cross-sections by varying the velocity.  Wind velocity and shell
velocity are assumed to be the same, allowing us to use an analytical
description, rather than a numerical model, for the shell morphology.

The second to last column in Table~\ref{tab:sim} gives the efficiency
of converting shock kinetic energy into radiated luminosity as found
in our simulations, based on the input kinetic energy and the
integrated luminosity in the light curve.  This is the maximum
efficiency corresponding to the bolometric luminosity output.  The
efficiency in converting shock kinetic energy to {\it visual} light
must be comparable to or less than this value.

Finally, Table~\ref{tab:sim} shows the the velocity of the SN remnant
after it has collided with the shell. We measure this velocity at a
fixed point in time, except where indicated otherwise; these
exceptions are necessary due to the nature of the circumstellar
medium, which may require a longer time interval before the supernova
has broken through the shell. Also, we don't list a final velocity for
those SNe that interact with bipolar shells, since for thise
simulations the velocity is angle-dependent.

\section{Numerical method}

We use the ZEUS~3D code \citep{SN92, c96} for our simulations.  The
grid is spherical and two-dimensional, with 500 gridcells along the
radial axis and 100 gridcells along the asimuthal axis, covering a 90
degree angle from pole to equator.  We have seeded both the initial
supernova and the circumstellar nebula with small scale density
fluctuations (5$\%$ for the supernova ejecta and $1\%$ for the
circumstellar medium).  This ensures that the supernova breaks up the
circumstellar shell upon collision.

\subsection{Grid evolution}

In order to achieve a high resolution at the collision between
supernova and circumstellar medium, the size of the radial gridcells
decreases with the radius.  This gives us the highest resolution at
the outer boundary.  Since we need to maintain this high resolution at
the collision front, we use the moving grid option that is part of the
ZEUS 3D code (see \citealt{wetal08, vVetal09}).  At the start of the
simulation the freely expanding supernova fills the entire grid, with
the exception of the outer radial boundary, which is set to an inflow
boundary condition with the parameters of the circumstellar medium
that the supernova is running into.  At the end of each timestep the
code finds the highest radial velocity within 50 radial gridcells of
the outer boundary.  Using the velocity in this cell as a basis all
gridcells are moved outward as well, with velocities:

\begin{equation}
v_{\rm grid}[i]~=~ 2 v[ic] \frac{r[i] - r[0]}{r[ic]-r[0]},
\end{equation}

\noindent with $r[i]$ the radius of the gridcell with index $i$, which
runs from 0 to 500, $ic$ the index of the gridcell in which the radial
velocity is highest and $v[ic]$ the highest radial velocity within 50
gridcells of the outer boundary.  The physical conditions at the outer
boundary are updated each time the grid expands to conform to the
values of the circumstellar medium at that particular radius.  In this
way the entire grid is stretched in the radial direction, ensuring
that a) the supernova remnant can never overrun the outer boundary; b)
a high resolution is always maintained close to the outer boundary
where the collision takes place and c) the inner boundary is fixed and
does not move.  (N.B. This method works well as long as one has to
deal with a strong shock.  It is not recommended for subsonic
expansion).

A drawback of this method is that the circumstellar nebula is supposed
to be static during the SN expansion, whereas speeds of the pre-shock
CSM for luminous SNe~IIn seen in narrow P Cygni absorption features
tend to be of order 100--500 km s$^{-1}$
\citep{S07,S08a,S09b,Tetal08}.  (The CSM speeds listed in Table 1
essentially determine the radii of the shells and therefore their
density for an assumed total mass.)  However, the velocities of the SN
ejecta expansion are much faster than those in the CSM nebula, such
that any evolution of the nebula during the SN expansion phase can be
considered small.  The inner radial boundary is fixed at $r=0$ and
does not move when the grid expands.  The inner radial boundary and
both azimuthal boundaries are set to reflecting boundary conditions so
no matter can escape from the system.

\subsection{Radiative cooling}

In order to obtain a lightcurve from our simulation we include the
effect of optically thin radiative cooling, using the cooling curve
from \citet{MB81}.  We extend this cooling curve to temperatures above
$10^{10}$~K by assuming that for these temperatures the cooling curve
depends on the temperature as $\Lambda(T)~\sim~\sqrt{T}$ (i.e.,
Brehmstrahlung).

Rather than use the cooling routine that comes as part of the ZEUS~3D
code, we implement a new numerical method, described by \cite{t09}.
This method uses exact integration of the radiative cooling function
rather than the traditional implicit or explicit schemes.  It is
faster, more accurate and avoids the potential instability of the old
radiative cooling method used in the ZEUS~3D code, which uses a
Newton-Raphson implicit calculation scheme.

\begin{figure*}
 \centering
\resizebox{\hsize}{!}{\includegraphics[width=0.95\textwidth,angle=-90]{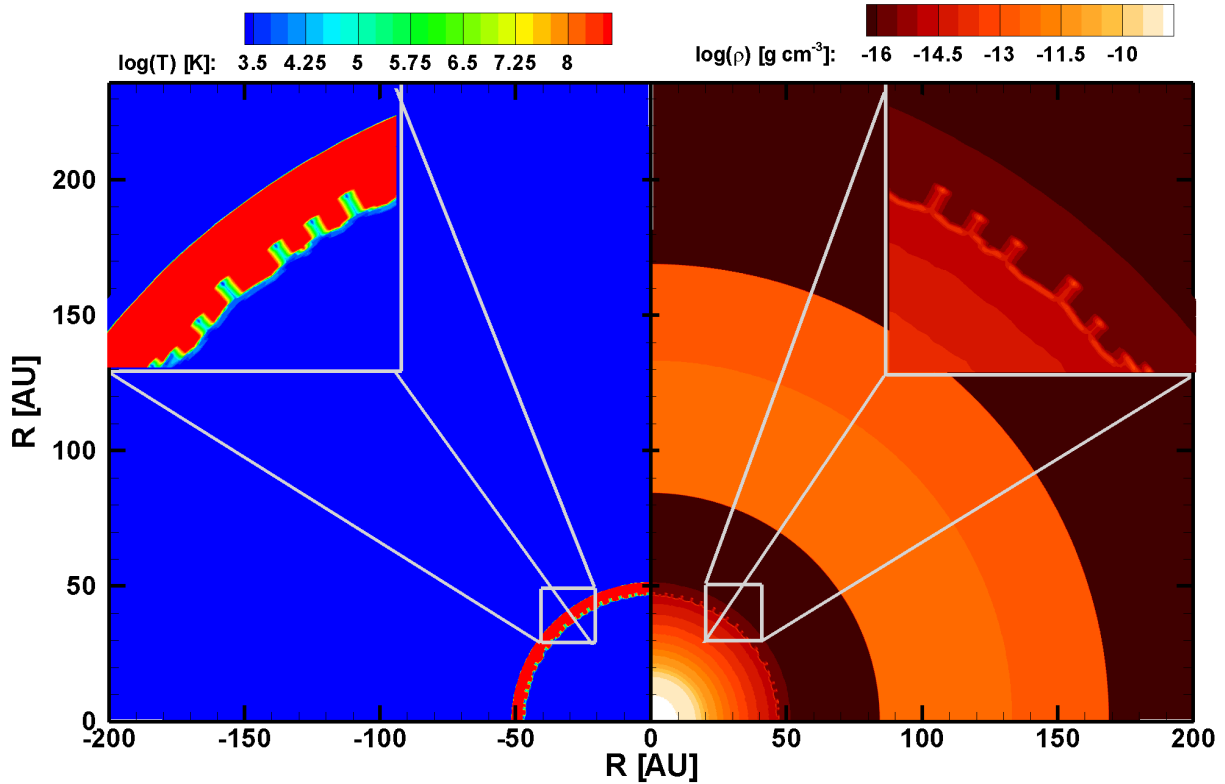}}
\caption{Temperature (left) and density (right) for simulation A03 at
  {t=11.5\, days} after the start of the simulation. The supernova has
  not yet collided with the circumstellar shell. The front of the
  supernova expansion (${\sim~50\, AU}$) is clearly visible because of
  the high (${\sim~10^8\, K}$) local temperature. The small insets show how instabilities form in the thin supernova shell. 
Clearly, we are at the limit of what can be achieved with this grid-resolution.}
 \label{fig:A03_10}
 \end{figure*}

\begin{figure*}
 \centering
\resizebox{\hsize}{!}{\includegraphics[width=0.95\textwidth,angle=-90]{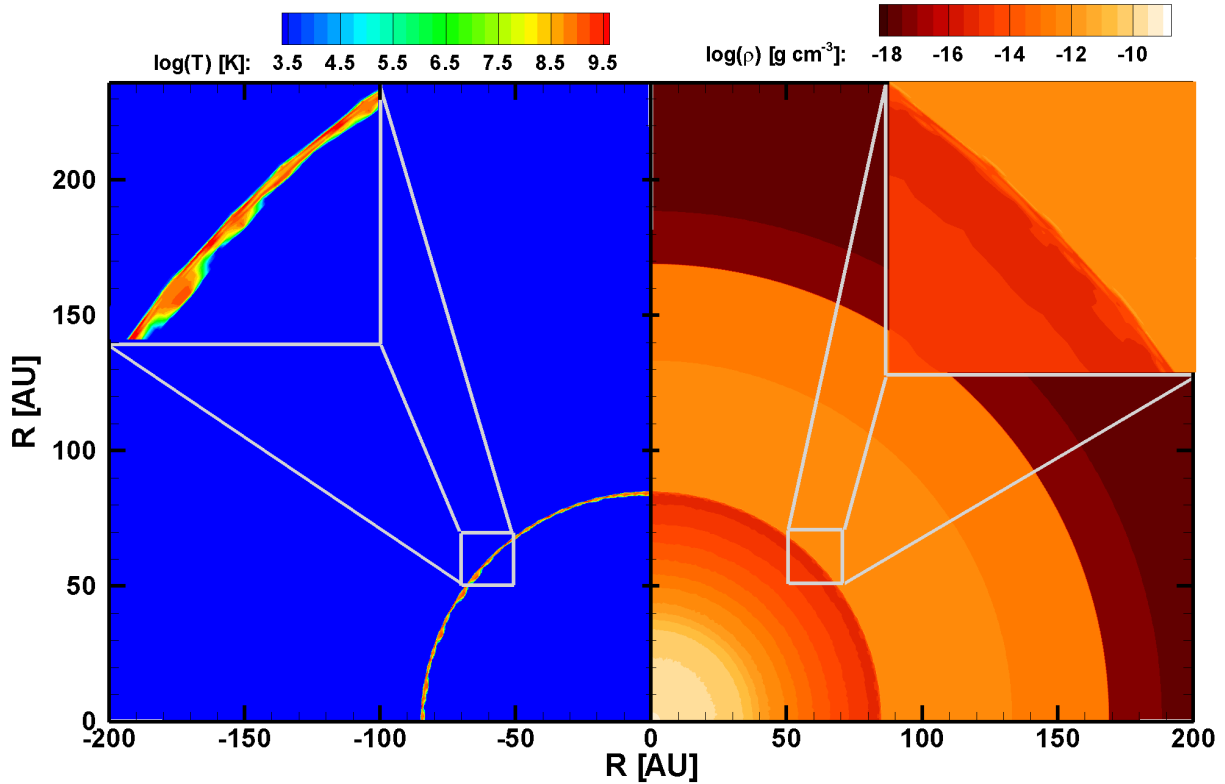}}
\caption{Similar to fig.~\ref{fig:A03_10}, but at {t=23\, days}. The
  supernova ejecta have reached the circumstellar shell.  Note that
  the high temperature region has become extremely narrow. This is due
  to the high density of the shocked gas, which allows it to cool very
  rapidly.  The temperature of the shocked gas increases, as more
  kinetic energy is converted to thermal energy. Again, the small figures show details of the supernova shell, which is extremely thin. 
Local instabilities are small.}
 \label{fig:A03_20}
\end{figure*}

\begin{figure*}
 \centering
\resizebox{\hsize}{!}{\includegraphics[width=0.95\textwidth,angle=-90]{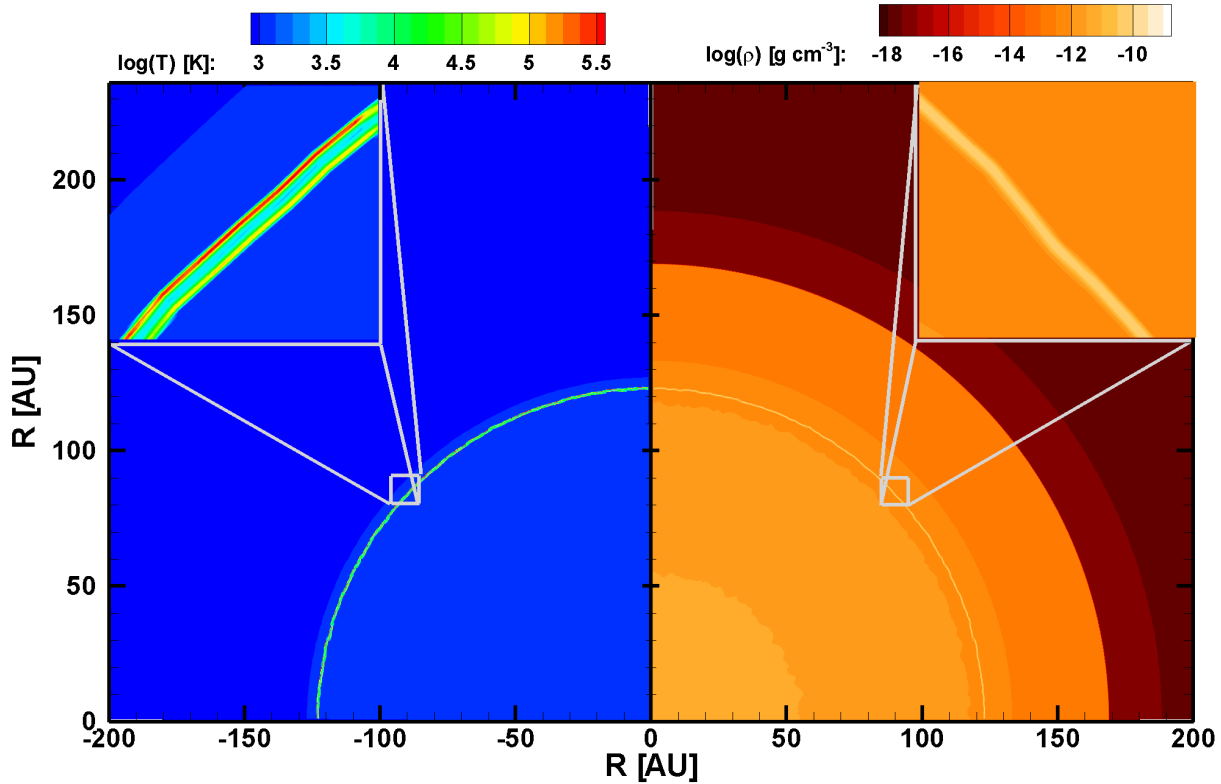}}
\caption{ Similar to fig.~\ref{fig:A03_10} to \ref{fig:A03_75} and ,
  but at t=86.8 days.  The supernova is approx. halfway through the
  circumstellar shell.  The temperature of the shocked region is much
  reduced (to ${\sim~10^5\, K}$), because the high density of the
  shell reduced the expansion velocity. The small figures show the thin high temperature layers on each side of the shell. The shell is not perfectly spherical, but the 
instabilities are extremely small as they are compressed between the expanding supernova and the high density material of the shell.}
 \label{fig:A03_75}
\end{figure*}

\begin{figure*}
 \centering
\resizebox{\hsize}{!}{\includegraphics[width=0.95\textwidth,angle=-90]{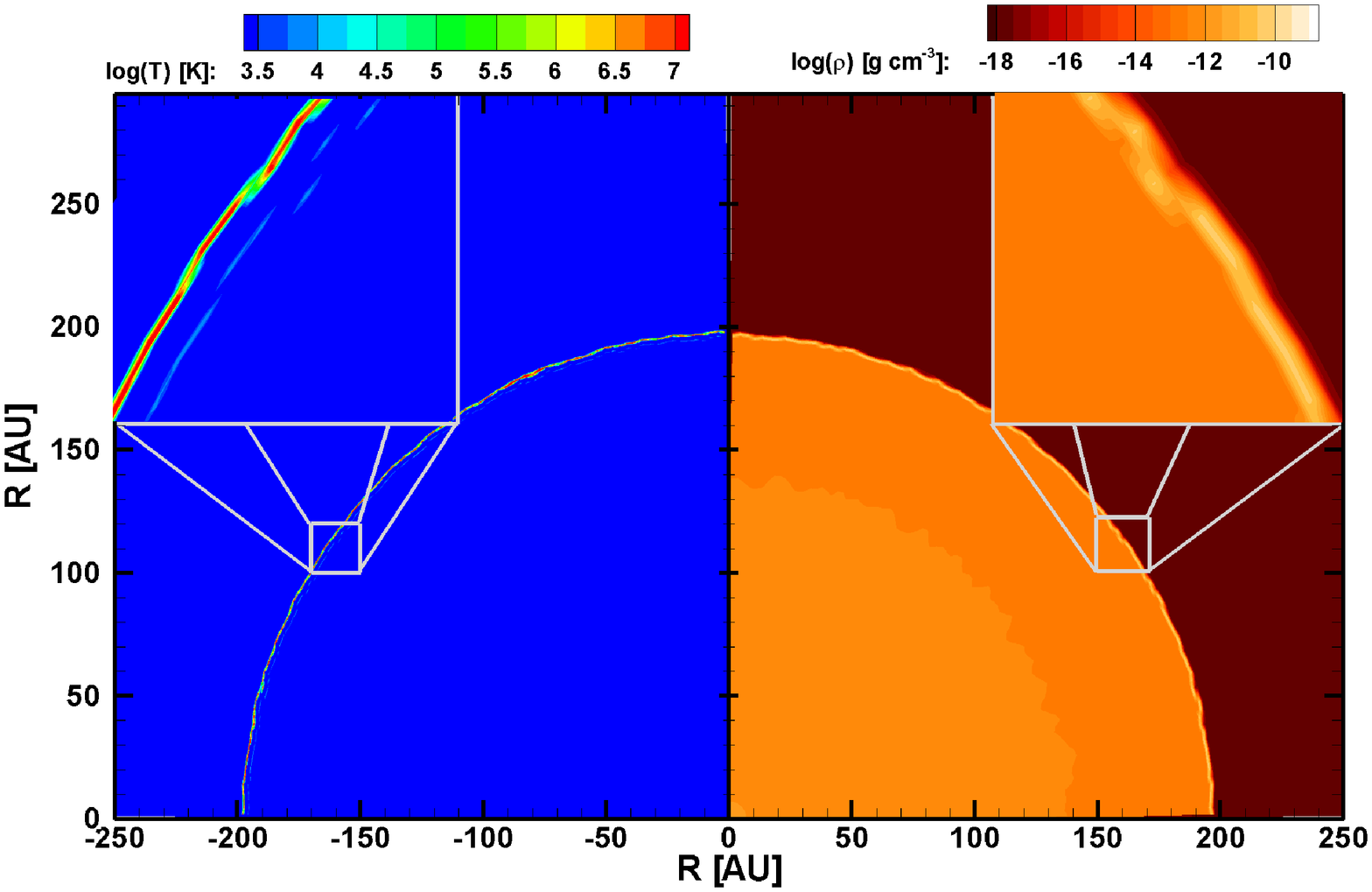}}
\caption{Similar to figs.~\ref{fig:A03_10} through \ref{fig:A03_75},
  but at t=173.6\, days.  The shock heated layer remains extremely
  thin, indicating a nearly isothermal shock.  The shock temperature
  has decreases because sweeping up the circumstellar matter slows
  down the supernova expansion. The instabilities are somewhat larger now, but remain small compared to the overal scale of the expansion.}
 \label{fig:A03_150}
\end{figure*}

\begin{figure*}
 \centering
\resizebox{\hsize}{!}{\includegraphics[width=0.95\textwidth,angle=-90]{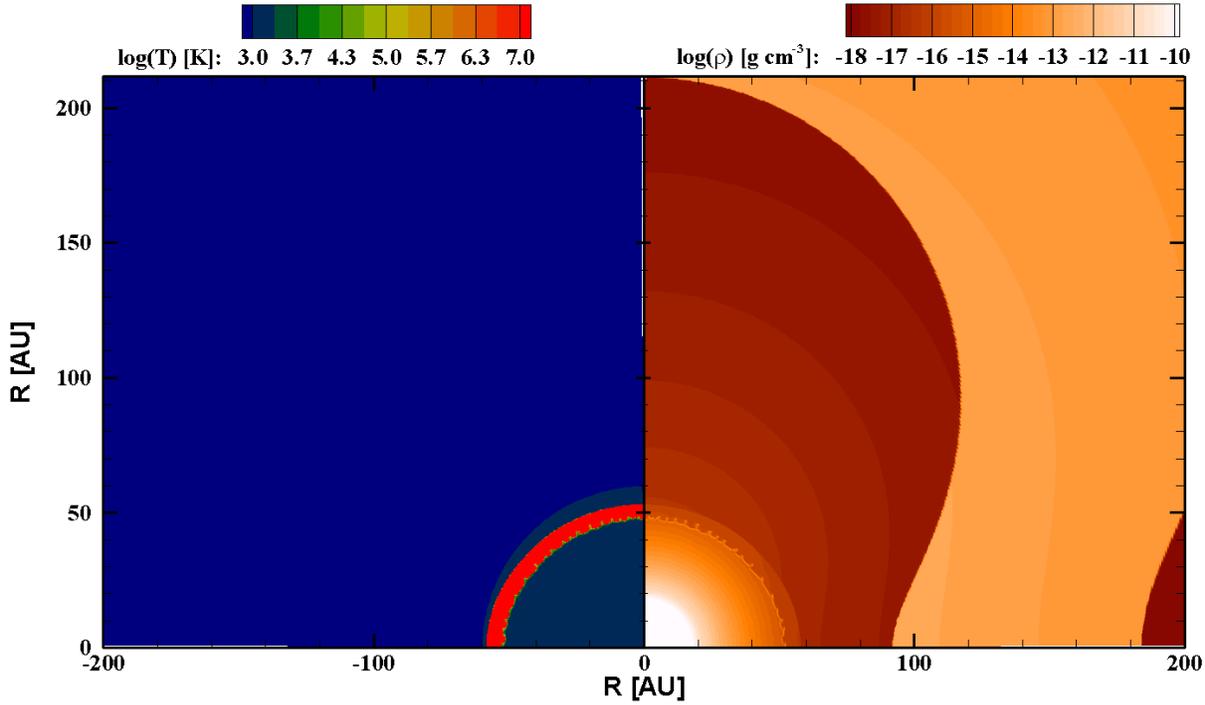}}
\caption{Temperature (left) and density (right) for simulation D02 at
  {t=11.5\, days} after the start of the simulation.  This is the
  equivalent of fig.~\ref{fig:A03_10}, but with a bipolar nebula.  At
  this point in time the supernova expansion is almost identical to
  the expansion in a spherical CSM.}
 \label{fig:D02_10}
 \end{figure*}

\begin{figure*}
 \centering
\resizebox{\hsize}{!}{\includegraphics[width=0.95\textwidth,angle=-90]{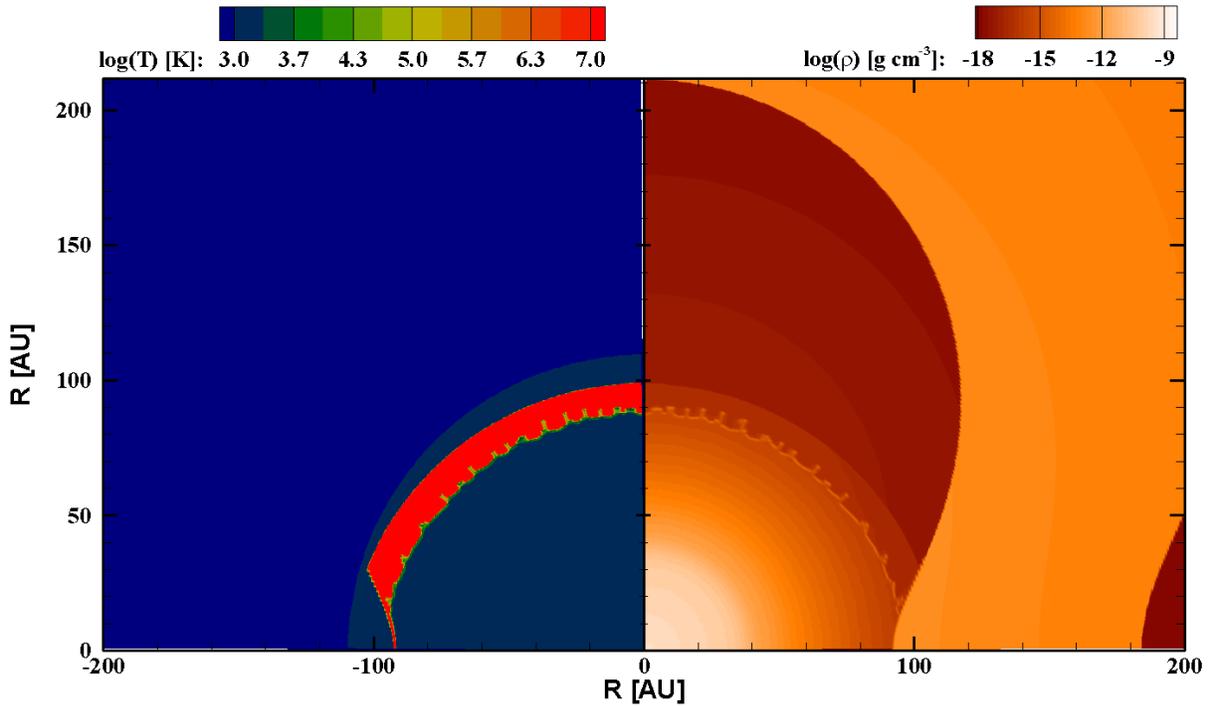}}
\caption{Similar to fig.~\ref{fig:A03_20} (same timestep), but for
  simulation D02. At the equator the supernova has reached the shell
  and has been slowed down abruptly at the pole the supernova is still
  expanding into the wind. Note the difference in the hot gas layer,
  which has been squeezed by the collision.}
 \label{fig:D02_20}
 \end{figure*}

\begin{figure*}
 \centering
\resizebox{\hsize}{!}{\includegraphics[width=0.95\textwidth,angle=-90]{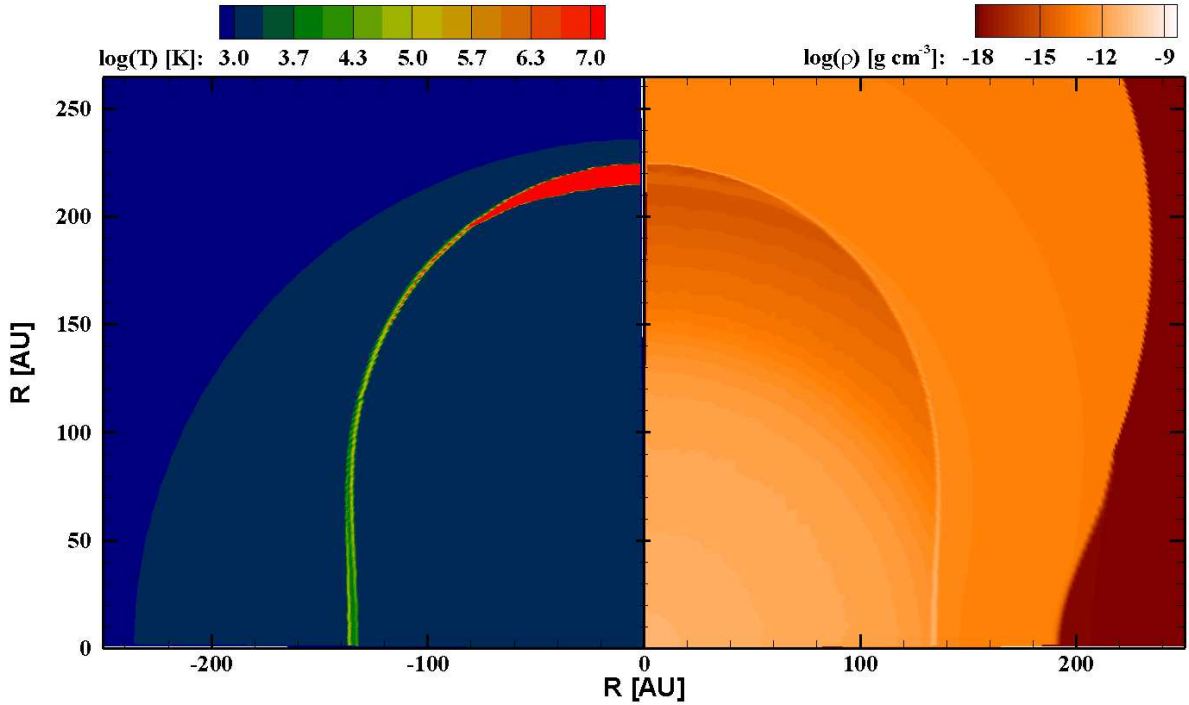}}
\caption{Similar to fig.~\ref{fig:A03_75} (same timestep), but for
  simulation D02.  At the equator the supernova is moving through the
  shell.  The slow shock has reduced the temperature of the hot gas
  zone to about $10^4$\, K.  At the pole the supernova has finally
  reached the circumstellar shell. There the hot gas is still at a
  high temperature ($10^7$\, K).  }
 \label{fig:D02_75}
 \end{figure*}

\begin{figure*}
 \centering
\resizebox{\hsize}{!}{\includegraphics[width=0.95\textwidth,angle=-90]{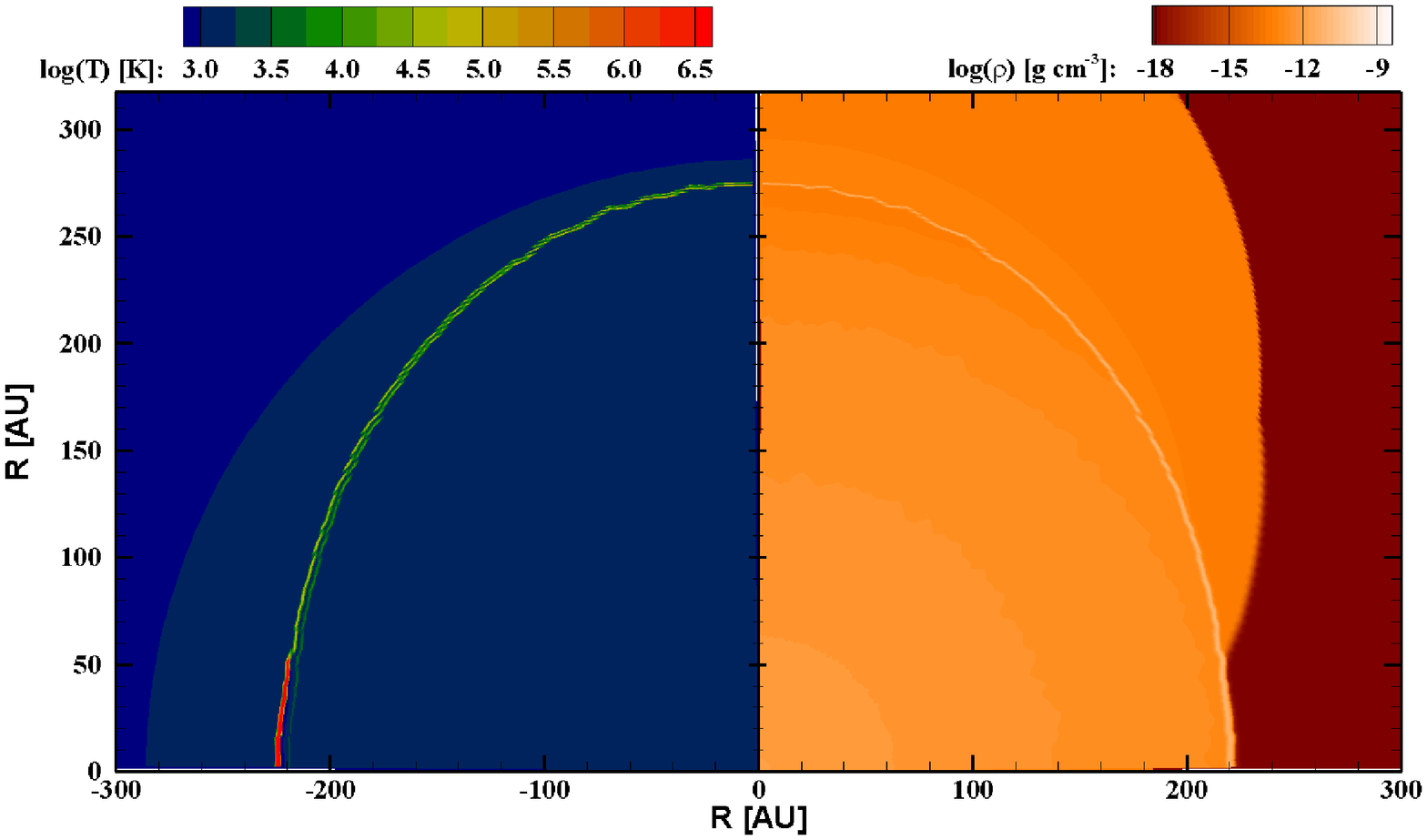}}
\caption{Similar to figs.~\ref{fig:A03_150} (same timestep), but for
  simulation D02.  The supernova has broken through the shell at the
  equator, but is still plowing through it at the pole.  The
  temperature at the equator is now high again ($\sim 10^6$\, K),
  whereas the shock over the poel has slowed down, lowering the local
  shocked gas temperature.}
 \label{fig:D02_150}
 \end{figure*}

The assumption of {\emph{optically thin}} radiative cooling to
  generate our lightcurves has some drawbacks.  The circumstellar
  shells used in our simulations have high densities and are therefore
  likely to be optically thick to Thomson scattering if fully ionized.
  However, at such high densities, it is difficult for the material to
  remain fully ionized because of fast recombination rates, we believe
  that our assumption is acceptable for our limited pruposes, at least
  as far as radiation in the optical part of the spectrum is
  concerned.  The high density of the circumstellar shells makes it
  unlikely that ultraviolet the light from the SN itself can fully
  ionize them.  Those areas of the shell that become photo-ionized
  will undergo recombination on a very short timescale.  Typical mass
  density in the shell is about $5\times
  10^{-13}~\mathrm{g}/\mathrm{cm}^{3}$ (see figs.~\ref{fig:A03_10}
  through \ref{fig:D02_150}).  Assuming pure ionized hydrogen for the
  sake of simplicity this gives us an electron density $n_e$ of
  $6\times10^{11}~\mathrm{cm}^{-3}$.  \citet{DW} give of recombination
  rate of

\begin{equation}
\dot{N}_R~=~n_e^2 \beta_2(T_e) 
\end{equation}

\noindent with $\beta_2(T_e)~=~2\times
10^{-10}T_e^{-3/4}~\mathrm{cm}^{3}/s$.  For an electron temperature
$T_e$ of 10\,000~K, this gives us a recombination rate of
$7.2\times10^{10}$ per second.  So even if fully ionized initially,
the shell will recombine very quickly compared to the expansion of
theshell.  It takes the SN at least several days to reach the inner
edge of the shell, so the effects of the initial ionization will most
likely have disappeared by then, leaving only the remaining radiation
from the expanding shock to photoionize the shell.  This greatly
reduces the number of free electrons that are available for
scattering.  Furthermore, although we use a shell with a smooth
density structure (apart from the small random variations mentioned
above), in reality circumstellar shells show a far more complicated
structure of high density filaments interspaced between low density
areas.  Under these circumstances, the photons will tend to escape
through the low density regions \citep{OGS04,OC07}.  Finally,
scattering by itself does not necessarily change the shape of the
emerging lightcurve since a photon can escape with little modification
even after multiple scatterings.  Therefore, even though the electron
scattering optical depth of our denser shells (under assmption of full
ionization), can be as large as $\tau_e \geq 100$, the true optical
depth will be much smaller due to efficient recombination.

The shape of the lightcurve will change if the diffusion time for
photons to escape from the circumstellar shell gets close to the
actual expansion time of the SN \citep{SM07}.  However, this is only
likely to affect the light curves at early times; the net effect would
be a slower rise time to peak luminosity and possibly a smoother peak,
while diffusion is unlikely to substantially affect the overall
efficiency of converting kinetic energy into radiation.  The typical
diffusion timescale of a photon through the shell is $t_{\mathrm diff}
= \tau D/c$, with $\tau$ the optical depth, $D$ the cross-section of
the shell and $c$ the speed of light, whereas the expansion velocity
is $t_{\mathrm exp} = D/V$.  With the expansion velocity $V$ typically
below 2000~$\kms$ (see the shell velocity plotted in
fig.~\ref{fig:V_A03} and also typical final velocities in
Table~\ref{tab:sim}) and lower for the denser, more optically thick
shells, the photons have time to escape from the shell ahead of the
expanding supernova.

Although we have attempted to account for radiative cooling in a
realistic way in our calculations, this is a difficult problem and our
method is simplified and necessarily limited.  Therefore, when
interpreting our results, we concentrate on relative changes from one
model to the next as we vary input parameters like mass and speed,
rather than the absolute values of the luminosity for any individual
model.  As noted earlier, it is not our goal here to fit the observed
lightcurve and derive corresponding physical parameters for any
individual SN, but rather, we aim to understand how the variety of
possible observed properties arises from different input parameters.

\section{SN-CSM interaction} 
\label{sec-sn-csm}

Here we describe the general properties of our SN-CSM interaction
simulations.  As a SN interacts with the CSM, we observe three phases
dictated by our assumed input geometry: (1) A fast interaction between
the SN and the (relatively) low-density wind inside the shell, (2) a
slower interaction as the SN shock pushes into the much denser medium
of the massive shell, and (3) the final expansion phase as the SN has
broken through the shell and continues to interact with the wind
outside the dense shell.

To demonstrate the strong interaction between a core-collapse SN and a
circumstellar shell, consider simulation A03 (Table~\ref{tab:sim}), in
which a 30$\mso$ explosion collides with a 10$\mso$ circumstellar
shell moving at 200$\kms$.  Figures~\ref{fig:A03_10} to
\ref{fig:A03_150} show snapshots of the temperature and density of the
expanding SN as it interacts with the CSM (movies of our simulations
are provided in the electronic edition).  The high post-shock density
encountered because of the very massive CSM shells we use causes the
radiative cooling to be extremely efficient in these simulations,
sometimes reducing the internal energy of even the shocked gas to the
point where the temperature reaches a minimum value of
1,000~K.\footnote{The fact that our calculations cool to a temperature
  as low as 1,000~K with a standard cooling prescription has
  far-reaching implications for understanding dust formation in
  CSM-interaction SNe.  This is not the topic of our study here, but
  recent observations of SN~2006jc \citep{S08b} and 2005ip
  \citep{S09a,F09} have demonstrated that new dust grains are seen to
  condense in the post-shock gas at the same time when strong X-rays
  and high ionization emission lines are seen. With efficient cooling
  in the dense shock leading to the low temperatures in our
  simulations, dust formation may be a natural consequence.}  (This
lower limit is a matter of numerical convenience that we impose upon
the calculation.)  Since the temperature difference before and after
cooling can be quite large, we show the temperature of the gas before
the radiative cooling has been taken into account, which is more
representative for the wavelength of the emitted radiation.  This is
the same temperature that we use to obtain the lightcurves in
\S~\ref{sec-lcurve}, adjusted for adiabatic expansion.

At first, the SN ejecta expand quickly as the forward shock encounters
the stellar wind, creating a layer of hot (several times $\sim~10^8\,
K$), shocked gas (fig.~\ref{fig:A03_10}).  At the inner boundary of
this high temperature zone (the reverse shock), SN material piles up
and creates a shell.  The interaction is (nearly) energy conserving at
this point.  When gas crosses the reverse shock, the kinetic energy of
the SN is converted to internal energy and heats the shocked gas,
pushing the forward shock into the CSM.  The shell, which is very
  thin due to radiative cooling, is subject to thin-shell
  instabilities. However, these take time to form and the shell is
  expanding rapidly, which limits their opportunity to grow.  As a
  result, the shell retains its basically spherical shape.  Because
only a small fraction of the SN material has a high velocity (see
\S~\ref{sec-snmodel}), the blast wave slows down quickly when it
sweeps up the wind.  This effect is greatly increased in the next
phase when the SN ejecta collide with the dense circumstellar shell.

Initially, the collision between SN ejecta and a massive circumstellar
shell causes a rapid decrease of the forward shock velocity.  This
deceleration drains energy from the forward shock, \emph{and powers
  the main peak of the light curve}.  The density at the forward shock
increases sharply as the shock overtakes more of the massive
circumstellar shell.  The layer of hot, shocked gas is compressed as
the reverse shock starts to overtake the forward shock, which leads to
an increase in local temperature (${\sim~10^9\, K}$ in
fig.~\ref{fig:A03_20}).  The high temperature, combined with the high
density of the gas makes the radiative cooling efficient.  Therefore
the thermal pressure of the shocked gas does not increase further.
This, combined with the compression between the two shocks causes the
hot gas layer to become quite thin and marks the transition from an
energy conserving shock to a momentum conserving one.

As the forward motion slows, the shock temperature decreases.  The
cooling remains efficient, so the high temperature region, which is
now at about ${\lesssim~10^5\, K}$, remains thin (see
fig.~\ref{fig:A03_75}).  The thin shocked gas layer is subject to
  radiative cooling instability (the higher density regions cool more
  efficiently, leading to a loss of thermal pressure, which in turn
  leads them to be compressed to even higher density).  This can be
  observed in fig.~\ref{fig:A03_75} as variations in the local
  temperature in the shocked gas.  However, like in the initial phase,
  the expansion of the SN occurs at higher velocity than the formation
  of the instabilities.  Also, the shocked gas layer is compressed
  between two areas with very high density (the shell on the outside
  and the rest of the supernova on the inside), which inhibits
  expansion apart from the bulk motion of the shock.  Therefore, there
  is no significant departure from spherical symmetry.

Once the SN breaks through the shell, the forward shock may accelerate
again due to the transition to much lower densities in the wind,
though it will never reach the original high velocity because a large
amount of energy has been lost to radiation during the shell collision
phase.  Also, the velocity of the unshocked SN ejecta piling up at the
reverse shock decreases over time, limiting the shell's ability to
accelerate in this later phase.  As a result, the temperature of the
hottest shocked gas is now limited to a few times $10^6\, K$
(fig.~\ref{fig:A03_150}).  The lower density at the interaction front
makes the radiative cooling less efficient, which allows the hot gas
layer to build up, though it never reaches its original size.

\subsection{Bipolar nebulae} \label{sec-bipol1}

The collision between a SN and a bipolar nebula shows the same general
pattern as described above, but is somewhat more complicated and
modified by the shell geometry.  Figures~\ref{fig:D02_10} through
\ref{fig:D02_150} show the same time frames for simulation D02, which
models a collision between the same SN as in simulation A03 with a
10$\mso$ shell, but here the shell is bipolar.

Initially (fig.~\ref{fig:D02_10}) the simulations look the same as
before, but they diverge once the SN hits the circumstellar shell.
This occurs first at the pinched equatorial waist of the nebula, where
the shell radius is smallest (fig.~\ref{fig:D02_20}).  The collision
squeezes the region of shocked gas into a very thin layer.  In polar
directions, the SN still expands into a lower-density wind and the hot
gas layer remains wide.  At a later stage, the interaction with the
shell has slowed the expansion at the equator, leading to a lower
shock temperature ($\sim$10$^4$ K in fig.~\ref{fig:D02_75}), whereas
the shock temperature at the pole is still high because the shock has
only just reached the circumstellar shell and has not swept-up enough
mass to decelerate yet.

Eventually, the SN will start to break out of the shell at the equator
first, while it is still inside the shell at the pole (see
fig.~\ref{fig:D02_150}).  When this happens, the shock at the equator
will reheat to about $10^6$~K, while the temperature at the pole
remains low.  Since the circumstellar shell has most of its mass
concentrated at the pole (where it also has the largest solid angle),
it takes much longer for the SN to break out in that direction.  As a
result, the shock in the polar direction will always be less energetic
afterwards than at the equator.  Because of the different times when
the shock hits the equatorial and polar regions of the shell,
different shock temperatures can be seen simultaneously.  We therefore
suggest that simultaneous observations of multiwavelength (i.e. X-ray
and visual) light curves may provide a way to distinguish bipolar from
spherical shells, as we describe in more detail later.

\section{Supernova lightcurves}
\label{sec-lcurve}

\subsection{General Properties}

The assumption of optically thin cooling, though a reasonable
  approximation in optical wavelengths, breaks down for high
  frequencies.  Most likely, for massive shell collisions, the
  early-time X-rays and UV would be completely self-absorbed and
  reprocessed into visual-wavelength luminosity.  Therefore, rather
  than attempt to plot the emission as a function of the gas
  temperature, we concentrate on the bolometric luminosity light
  curves as a likely proxy for the visual lightcurves in later
  sections; this assumption may break down at late times when the
  shock becomes optically thin and X-rays can escape (see below).  In
  addition, it is important to note that our light curves correspond
  only to radiative energy losses from the post-shock gas.  We do not
  include the photospheric emission from the underlying SN itself,
  which could, in principle, be any type of SN.  (It is the shell
  collision that leads to a Type IIn spectrum and the enhanced
  luminosity, rather than any intrinsic property of the SN.)

As is shown below, the overall shape of the lightcurve for any
  SN-CSM collision model has the same general properties.  Initially,
  the SN expands into the (relatively) low-density wind, starting at
  high luminosity due to its high velocity.  As the expanding shock
  sweeps up more wind material, the expansion speed is reduced and the
  lightcurve shows a corresponding decrease in luminosity.

Note that the behavior in this early phase --- while the shock
propagates through the wind on its way to reach the inner radius of
the dense shell --- depends strongly on our assumed value for the
inner radius of the shell and on the assumed time before the SN when
the shell ejection finished ($t_{end}$ in Table 1).  If the SN had
occurred immediately after the shell ejection stopped or while it was
still in progress, then this early phase would not exist.  This may be
an important consideration in determining the early light curve shape:
some luminous SNe like SN~2006gy and SN~2005gj show a long and slow
rise to peak luminosity \citep{S07,P07}, while others are discovered
at peak and decline immediately, as in the cases of SNe~1998S, 1997cy,
and 2006tf \citep{L02,G00,S08a}, suggesting a very rapid initial rise
time.

When the supernova reaches the circumstellar shell, which takes
  on the order of 10-25 days in most of our simulations, the expansion
  decelerates abruptly.  This shows up as a rapid increase of the
  emission, because the fraction of kinetic energy converted into
  thermal energy is now high Also, the very high density of the
  shocked gas causes it to radiate very efficiently.  As the SN plows
  through the shell, the emission decreases again due to the general
  decrease in shock velocity, but remains high compared to the
  emission from the initial phase. 

Once the SN has overtaken the massive shell and begins to expand into
the outer low-density wind (at $t\simeq$140 days), the total emission
decreases because the density of the gas that the SN collides with has
decreased.  Unlike the previous phases while the blast wave was
expanding inside and through the massive shell, the densities are
relatively low, and so optical depth effects are less likely to cause
complete self-absorption of high energy photons.  Thus, once the blast
wave has broken through the outer boundary of a hypothetical massive
shell, we would predict that soft X-ray emission could in fact be
observed.  Mass-loss rates derived from observations of this X-ray
emission would trace the normal wind mass-loss rate of the progenitor
star in the years \emph{before} it ejected the massive dense shell
that led to the enhanced optical luminosity; meanwhile, the optical
luminosity is still being emitted by teh dense shell.  Therefore, one
would not necessarily expect agreement in mass-loss rates derived from
observed optical and X-ray luminosities (see e.g., \citealt{S07}).  As
the SN blast wave continues to expand into the wind, it gradually
decelerates because the amount of swept-up gas increases over time.
This leads to a steady and slow reduction in total emission in the
years after the initial collision.

Since both the circumstellar shell and the SN are spherically
symmetric, the collision happens at the same moment everywhere.
Similarly, the SN will break through the shell at the same time all
around its circumference.  As a result, both sides of the main peak in
the lightcurve have very steep slopes, and the change in X-ray
emission would most likely be quite sudden.  This is partly a result
of our prescribed geometry of the shell, with a clean inner and outer
boundary.  Real circumstellar shells can show a wide variety of
different geometries, including multiple shells and high degrees of
clumping, which can vastly change the appearance of the lightcurves.
As one example, we explore the influence that a bipolar shape has on
the emergent light curve.  Our point here is not to provide an
exhaustive grid of simulations of possible light curves, but to simply
illustrate the behavior as we vary the parameters of the collision in
order to guide the interpretation of lightcurves of luminous SNe.  The
responses of the light curves to various parameters of the wind and
shell are described in the following sections.

The most important consequence of the SN-shell collision is that SN
kinetic energy is converted to thermal energy and then lost to
radiation.  The efficiency of this conversion is a key parameter for
interpreting the energy budgets of SNe IIn.  For each simulation
discussed below, we list the total efficiency in converting kinetic
energy into radiated energy over the course of the simulation,
$E_{rad}/E_{SN}$, in the second to final column of Table 1.  We find a
large range in the conversion efficiency, depending on the mass of the
shell as well as the mass of the SN.  For a circumstellar shell mass
of 10~$M_{\odot}$, the efficiency is typically 15--30\%.  Efficiency
increases with increasing density of the circumstellar shell (higher
shell mass, slower velocity, or both).  The efficiency also increases
for lighter SNe (higher ratios of $M_{shell}/M_{SN}$), because of
momentum conservation and the greater deceleration of the fast SN
ejecta.  We elaborate on these points for individual cases below.

\begin{figure}[t]
  \includegraphics[width=\columnwidth]{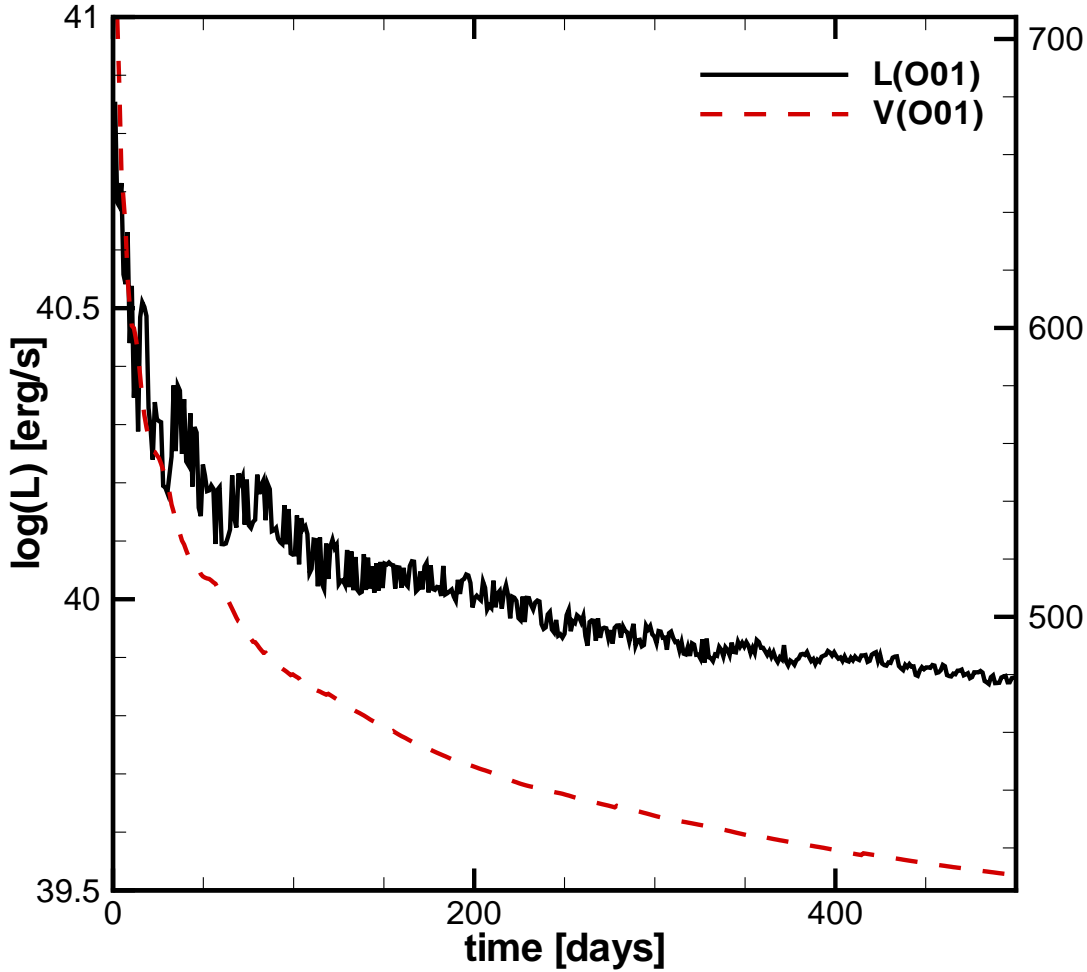}
  \caption{Lightcurve and reverse shock velocity for a SN expanding
    into a circumstellar medium that contains only wind (simulation
    O01).}
\label{fig:LV00}
\end{figure}

\begin{figure}[t]
  \includegraphics[width=\columnwidth]{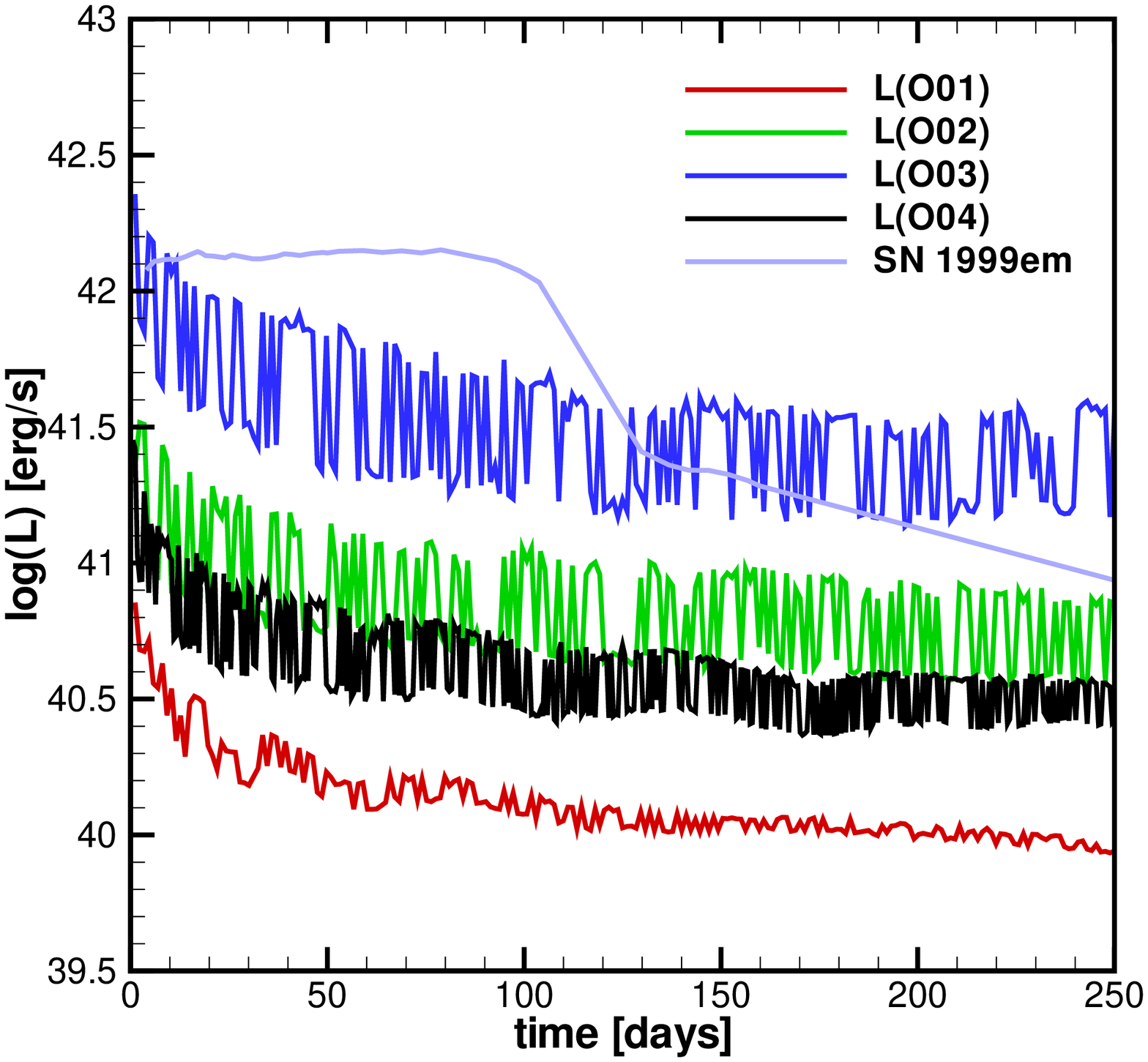}
  \caption{Lightcurves for four simulations without circumstellar
    shells. (O01-O04) In most cases the luminosity falls well below
    peak values for a typical SN~II-P photosphere
    ($\sim$10$^{42}$ergs), illustrated by the observed lightcurve of
    SN~1999em \citep{L02}.}
\label{fig:L_O0123}
\end{figure}

\subsection{No Shell, Just a Wind}

Since we are interested in investigating the effects of various
properties of massive circumstellar shells, one might first ask what
the collision looks like when there is no shell --- i.e., when it is
simply a collision between the SN and a dense steady
wind. Figure~\ref{fig:LV00} shows the bolometric luminosity emission
lightcurve and the shock velocity (See \S 6) for a simulation where
the circumstellar medium contains no shell (O01), but just a dense
wind with $\dot{M}=10^{-4}$ $M_{\odot}$ yr$^{-1}$ expanding at a speed
of 200 km s$^{-1}$ as one might expect for a massive luminous blue
variable (LBV) progenitor \citep{S07,S09b,Tetal08}.  Both the
luminosity and post-shock shell velocity start high, but decline
quickly as the SN sweeps into the dense wind.

However, an important point to take away from simulation O01 is that
the peak luminosity at early times is less than 10$^{41}$ ergs
s$^{-1}$, and is therefore likely to be dwarfed by much stronger
emission from an underlying SN photosphere (not shown in
Fig.~\ref{fig:LV00}).  A normal Type II-P supernova, such as SN~1999em
(Figure~\ref{fig:L_O0123}; \citealt{L02}), has a luminosity during a
$\sim$110 day plateau of $\sim$10$^{42}$ ergs s$^{-1}$.  This is
100$\times$ stronger than the day 100 luminosity in simulation O01.
Even SN~2005ip, which represents the lower end of luminosities for
Type IIn core collapse SNe, had a late-time luminosity due to
circumstellar interaction of 10$^{41.5}$ ergs s$^{-1}$
\citep{S09a,F09}.  The main consequence is that the more luminous
class of SNe~IIn require massive circumstellar shells, ejected in
outbursts occuring shortly before core collapse -- rather than steady
winds -- as emphasized elsewhere \citep{S07,S08a, SM07}.  This is also
illustrated by a comparison between simulation O01 and A00.  these
have the same input parameters except for a very low mass (0.1~$\mso$)
circumstellar shell in the case of A00. Despite the low mass, the
shell causes the total amount of energy converted to radiation to jump
by more than an order of magnitude.

Winds with higher density, either through high mass-loss rates (O02
and O03) or low velocity (O04) tend to produce higher luminosities
through the collision, as expected, but these enhancements are small
compared to the effect of massive shells.  (See also
Table~\ref{tab:sim} for the percentage of energy converted into
radiation.)  The only ``no-shell'' simulation to produce a higher
luminosity than that caused by even the smallest circumstellar shell
is simulation O03, which assumes a mass-loss rate of $10^{-2}\msoy$.
Interestingly, this high wind mass-loss rate produces a late-time
plateau with a luminosity of $\sim$10$^{41.5}$ ergs s$^{-1}$,
appropriate for the late phases of SN~2005ip \citep{S09a, F09}.  
  A similar progenitor mass-loss rate was inferred for SN~2005gl,
  which had an LBV-like progenitor identified in pre-explosion data
  \citep{GL09}. The shapes of these light curves also resemble
  SN~1988Z \citep{Aetal99}, where the luminosity remained high for
  about a decade, indicating that the expanding supernova interacted
  with an extended circumstellar wind, rather than a sharply confined
  shell.  Such a mass-loss rate is in excess of even the strongest
LBV winds in their quiescent states (i.e. 10$^{-3}\msoy$ in the case
of $\eta$ Car; \citet{H01}), but is comparable to smaller LBV
eruptions like the 1600 AD event of P~Cygni \citep{SH06} or the 1890
eruption of $\eta$ Car \citep{S05}.  In other words, a steady ``wind''
with $\dot{M} = 10^{-2} \msoy$ is essentially the same as a sustained
eruption (i.e.\ the total mass swept up by the shock is comparable).
This is also the only simulation without a shell for which the
radiative luminosity exceeds values typically expected from the SN
photosphere (Fig.~\ref{fig:L_O0123}).  One can expect that steady
winds or sustained eruptive phases with even higher mass-loss rates or
slower wind speeds will result in long-lasting light curves shaped
like those in Figure~\ref{fig:L_O0123}, but with even higher
luminosity.

%%% add lightcurve for A013 to this plot as well
\begin{figure}[t]
  \includegraphics[width=\columnwidth]{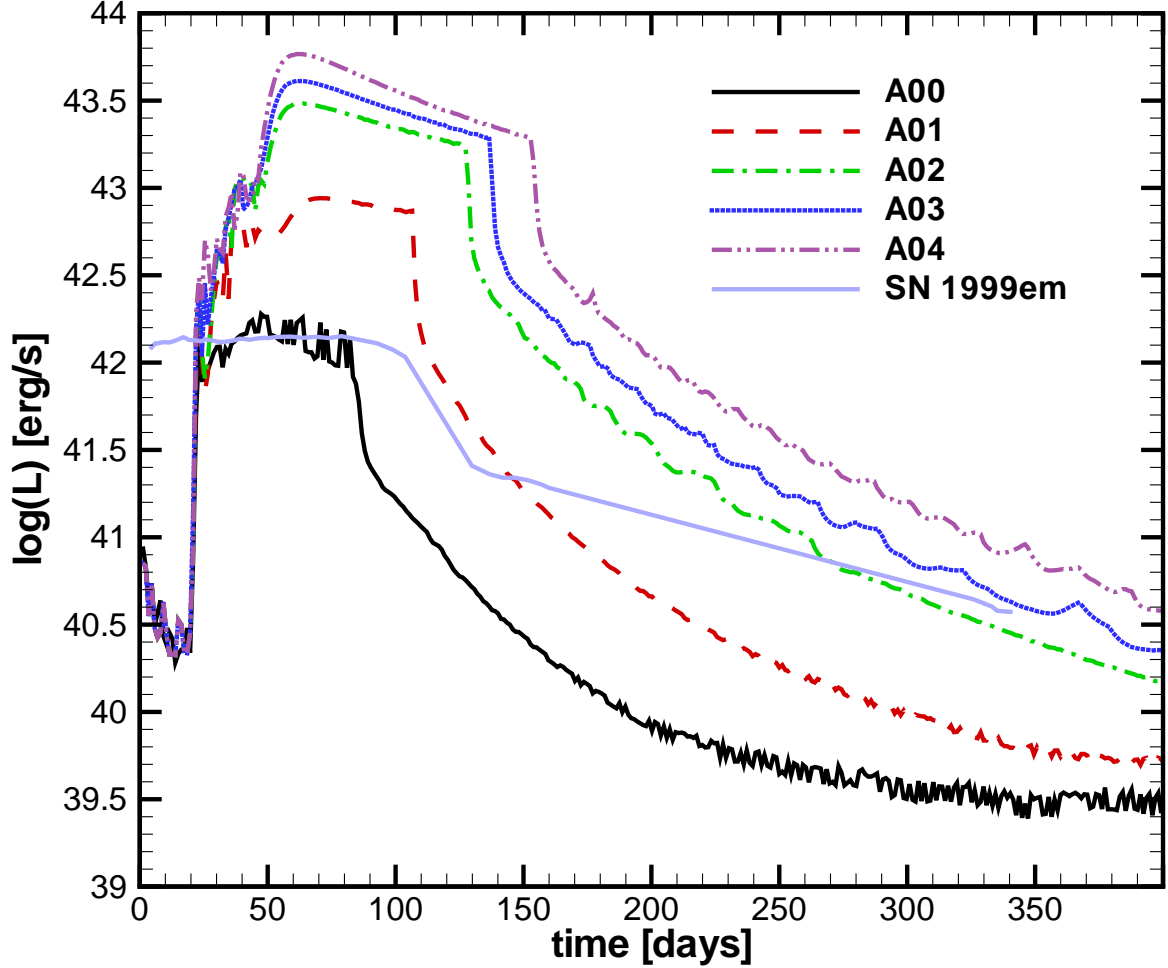}
  \caption{Total bolometric luminosities for simulations A00 through
    A04 as a function of time.  The higher shell masses cause higher
    luminosity peaks, for all other explosion and shell paramters held
    the same.  Higher shell masses also cause more deceleration, so
    the shock takes longer to break through the shell, leading to a
    longer lasting peak in the lightcurve.  Both the beginning and the
    end of the luminosity peak is marked by a sharp transition in all
    simulations, which results from our assumed inner and outer
    boundaries of the shell.  The lightcurve of the SN~II-P 1999em is
    shown again for comparison, as in Figure~\ref{fig:L_O0123}.}
\label{fig:L_mshell}
\end{figure}

\subsection{Shell masses}

The next group of simulations in Table~\ref{tab:sim} (A00 to A04)
explore the effect that the circumstellar shell mass has on the
evolution of the SN light curve.  Fig.~\ref{fig:L_mshell} shows the
total bolometric radiative luminosity for each simulation as a
function time, compared to the light curve of a normal
SN~II-P. Because it takes more energy to break through a more massive
shell, more kinetic energy is converted to thermal energy and then to
radiative energy loss.  Therefore, the higher the shell mass, the
higher the luminosity peak.  Also, it takes longer to break through a
high mass shell, because the shock suffers more deceleration, so the
duration of the peak luminosity will be longer for higher mass shells
as well.  It is noteworthy, that even the lowest mass shell ($0.1\
\mso$) simulation, A00, shows a clear peak and is therefore
distinguishable from the pure wind interactions shown in
fig.~\ref{fig:L_O0123}, although in practice this peak might be lost
amid the photospheric emission from the SN itself.

All shells show sharp transitions at the beginning and the end of the
main luminosity peak, but this is a direct result of our prescribed
sharp inner and outer boundaries of the shells.  It is a simplifying
assumption and is motivated by the observed sharp outer boundary in
some dense shells around massive stars, such as the Homunculus of
$\eta$ Carinae \citep{S06}, but it is not necessarily true in all
cases.  It is likely that some objects will have smoother density
transitions at the outer extent of the shell, and in those cases one
expects the CSM-interaction luminosity to drop more gradually.  The
plateau is almost horizontal for the lower mass shells in our study,
but changes to a shallow decrease with time for high mass shells.
This decrease results from the fact that the high mass shells
decelerate the blast wave to a greater extent as it plows through the
shell.  The decrease in shock speed leads to a reduction in post-shock
thermal energy and a lower emergent luminosity.

\begin{figure*}
 \centering
\resizebox{\hsize}{!}{\includegraphics[width=0.95\textwidth,angle=-90]{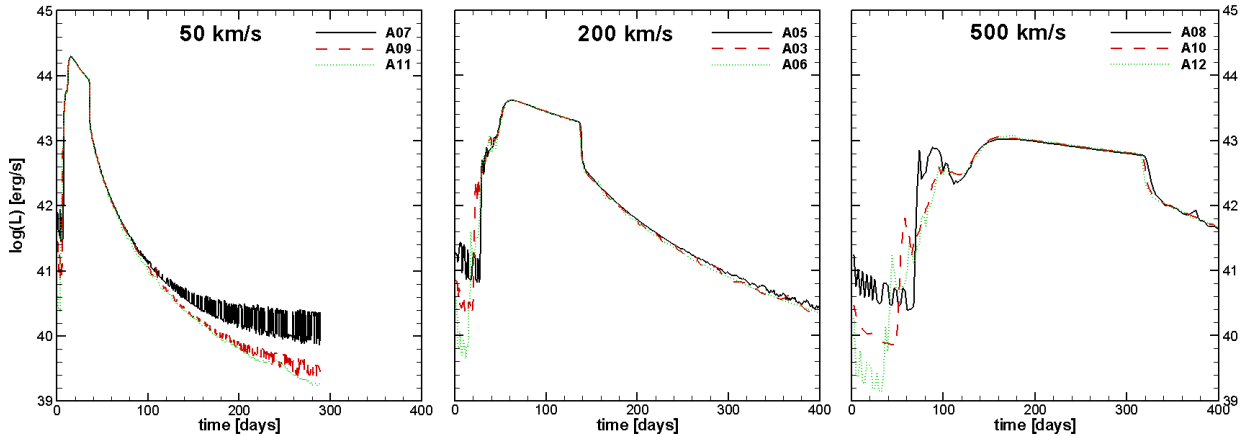}}      
\caption{Total radiative emission lightcurves for circumstellar matter
  with different velocities (50$\kms$ in the left panel, 200$\kms$ in
  the middle and 500$\kms$ on the right) and different wind mass loss
  rates (color coded lines). Obviously, the wind velocity has a major
  influence, since it determines how far the shell has travelled
  before the supernova hits it. Wind mass-loss rate makes very little
  difference, except in the very early stages.}
 \label{fig:L_v_mdot}
\end{figure*}

\subsection{Wind parameters}
\label{sec-wind}

In our simulations we vary both the wind velocity and mass-loss rate
to explore the influence of these parameters on the light curve.
Figure ~\ref{fig:L_v_mdot} shows the effect of the wind mass-loss rate
and velocity on the bolometric SN lightcurve, by comparing simulations
A07, A09 and A11 (left-hand panel in fig.~\ref{fig:L_v_mdot}), which
have identical parameters except for the wind mass-loss rate, which is
$10^{-3}$, $10^{-4}$ and $10^{-5}~\msoy$ respectively. In these
simulation the wind (and shell) velocity is fixed at $50\kms$.  In the
initial stage the difference is considerable, as the higher density
winds clearly create much stronger emission.  Also, the high wind
density in simulation A05 actually slows down the supernova expansion
more than the other two, delaying the moment when the expansion
reaches the shell, though not by a large amount.  Since the shells are
identical the lightcurves all have the same peak in the lightcurve.
After the circumstellar shell has been swept up, the difference
between the lightcurves is difficult to see.  The $10\mso$ shell slows
down the SN expansion to such an extent that the effect of the wind
mass-loss rate becomes negligible.  Still, after more time passes the
curves start to diverge, albeit slowly, with once again the highest
mass-loss rate creating the highest emission.

The middle and right-hand panels of fig.~\ref{fig:L_v_mdot} show the
same phenomena, but for wind (and shell) velocities of 200 and
$500\kms$ respectively.  The results follow the same pattern.
However, due to the higher velocties, the densities are generally
lower.  As a result, the influence of the wind mass-loss rates in the
final stages is lower for the simulations with wind velocity of
500$\kms$.

Comparing the three panels of fig.~\ref{fig:L_v_mdot} shows the effect
of wind velocity on the lightcurve.  Obviously, lower wind velocities
mean that the shell is closer to the star when the supernova hits is,
which means that the entire time-frame of the interaction gets
shortened.  Also, the density in the shell is higher
($\rho\,\sim\,1/v$), whereas the cross-section of the shell is
smaller, leading to a higher, narrower peak in the luminosity.

\begin{figure*}
 \centering
\resizebox{\hsize}{!}{\includegraphics[width=0.95\textwidth]{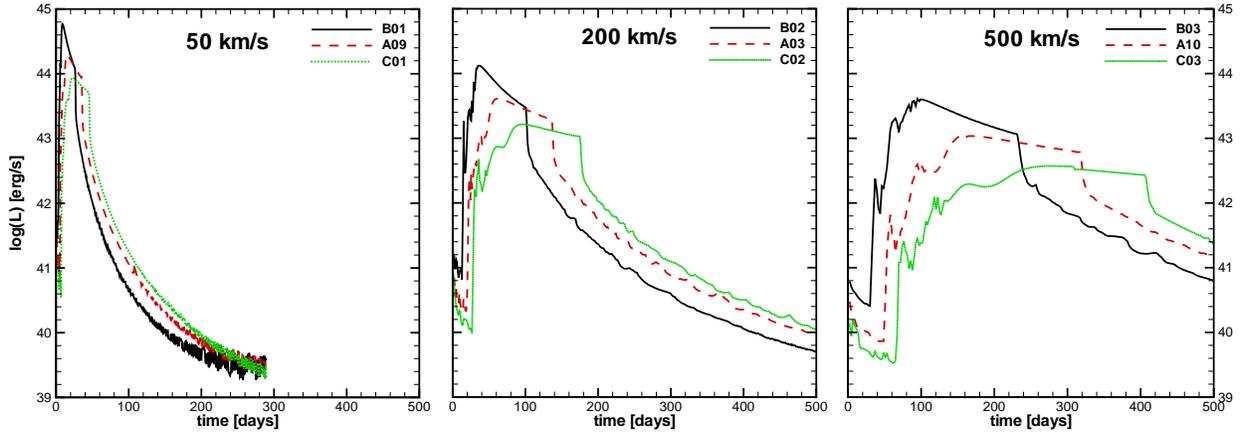}}  
\caption{Total radiative emission lightcurves for collsions with
  different supernova mass and wind velocities.  The higher the
  supernova mass, the lower the emission, as the high mass supernova
  has a relatively low velocity.}
 \label{fig:L_v_Msn}
\end{figure*}

\begin{figure}[t]
  \includegraphics[width=\columnwidth]{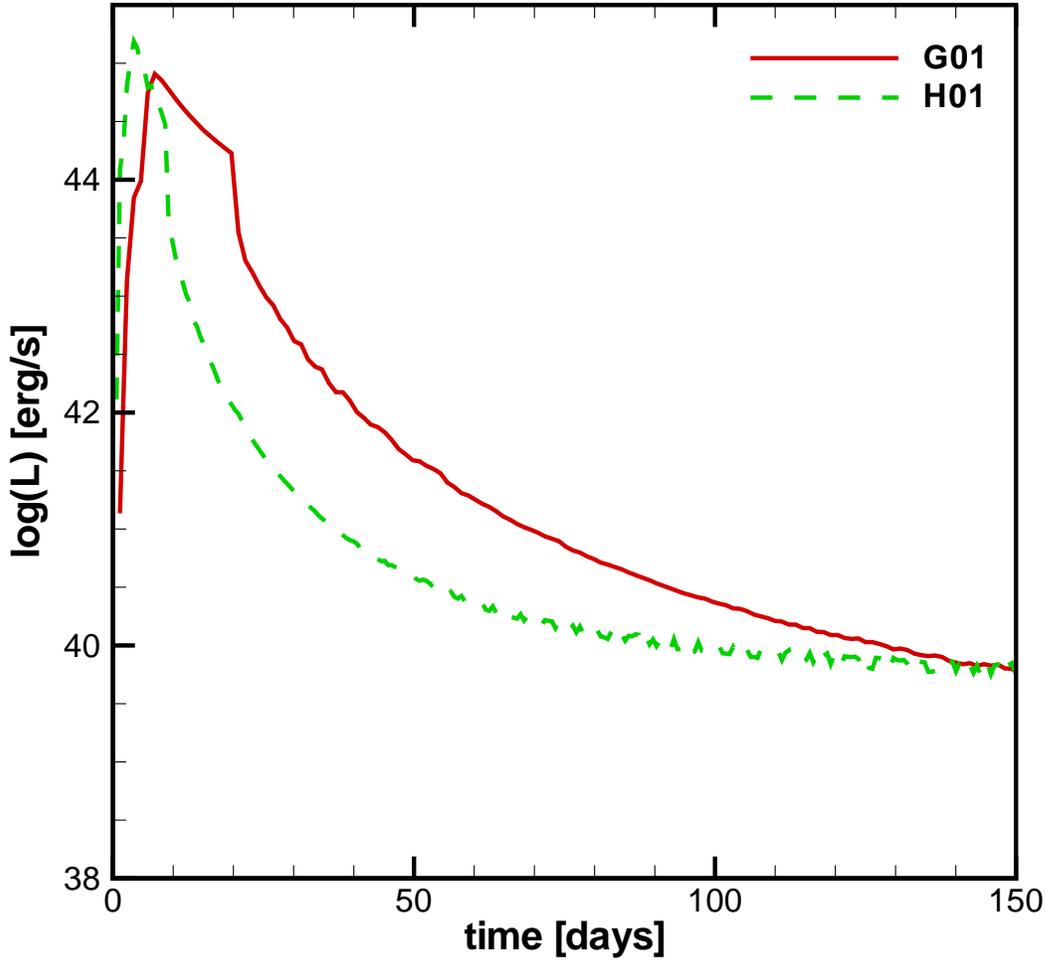}
  \caption{Lightcurves for two low mass supernovae (G01 and H01) with low mass shells. 
The collisions are extremely luminous, but fade quickly.}
\label{fig:L_G01H01}
\end{figure}

\subsection{Supernova masses}

Letting three different SNe interact with the same circumstellar shell
produces the light curves shown in Fig.\ref{fig:L_v_Msn}, which shows
the bolometric lightcurves for three different supernova masses (10,
30 and 60$\mso$: colorcoded lines), colliding with three different
circumstellar shells (velocities at 50, 200 and 500$\kms$: left,
center and righ respectively).  All three circumstellar shells have
the same mass of 10 $M_{\odot}$.

These lightcurves show two characteristic patterns: Because the
kinetic energy in the SN is the same for all three simulations, the
lower mass SNe have higher initial velocities. As a result, the peak
in luminosity that results from the collision between the SN expansion
and the circumstellar shells occurs earlier, and the peak luminosity
is higher because of the greater energy per unit mass that is lost to
radiation when the material is decelerated.  The low mass SNe have
less momentum ({$m_{A09}v_{A09}^2\,=\, m_{B01}v_{B01}^2$} and
{$m_{A09} \, =\, 3m_{B01}$} so {$m_{B01}v_{B01}\, = \,
  m_{A09}v_{A09}/\sqrt{3}$}), so they slow down and give up their
kinetic energy more quickly during the collision.  As a result, the
10~$\mso$ supernova produces a lightcurve where the flat plateau in
the lightcurve peak is sharply angled, rather than horizontal as for
the higher mass supernovae. The slope of this plateau may therefore
provide a useful diagnostic to constrain the mass and momentum ratios
of the underlying SN and CSM shell.  This same pattern can be seen in
all three figures.  The essential result is that relatively lower-mass
SNe (i.e. faster SNe) have higher efficiency in converting kinetic
energy into radiation, while more massive SNe have more momentum and
therefore lose less of their kinetic energy to radiation. This exactly
the opposite of the effect of the shell mass, which produces a higher
efficiency when the shell is more massive.  Therefore, the highest
efficiency will be achieved for those collisions wherein a relatively
low mass SN collides with a relatively high-mass shell. (See also
fig.~\ref{fig:efficiency}).

The influence of the wind and shell velocity is similar to that
observed in Fig.~\ref{fig:L_v_mdot}.  Higher expansion speeds stretch
out the duration of the light curve and lower the peak luminosity,
because the collision takes place later and over a longer time, and
the shock plows through a lower-density shell for the same shell mass.

Extreme cases of low SN mass can be seen in fig.~\ref{fig:L_G01H01},
which shows the lightcurves for simulations G01 and H01, where
supernovae of $6\mso$ and $1\mso$ respectively collide with shells of
equal mass.  The resulting lightcurves show peaks with extremely high
luminosity ($\sim$10$^{45}$ erg s$^{-1}$), comparable to those of the
most luminous SNe observed to date
\citep{O07,qetal07,metal2009,getal2009,S09b}.  However, due to the low
masses of the circumstellar shells, the bright peak fades quickly as
the shell is swept up within just a few weeks, which is faster than
the observed examples.

\begin{figure}[t]
  \includegraphics[width=\columnwidth]{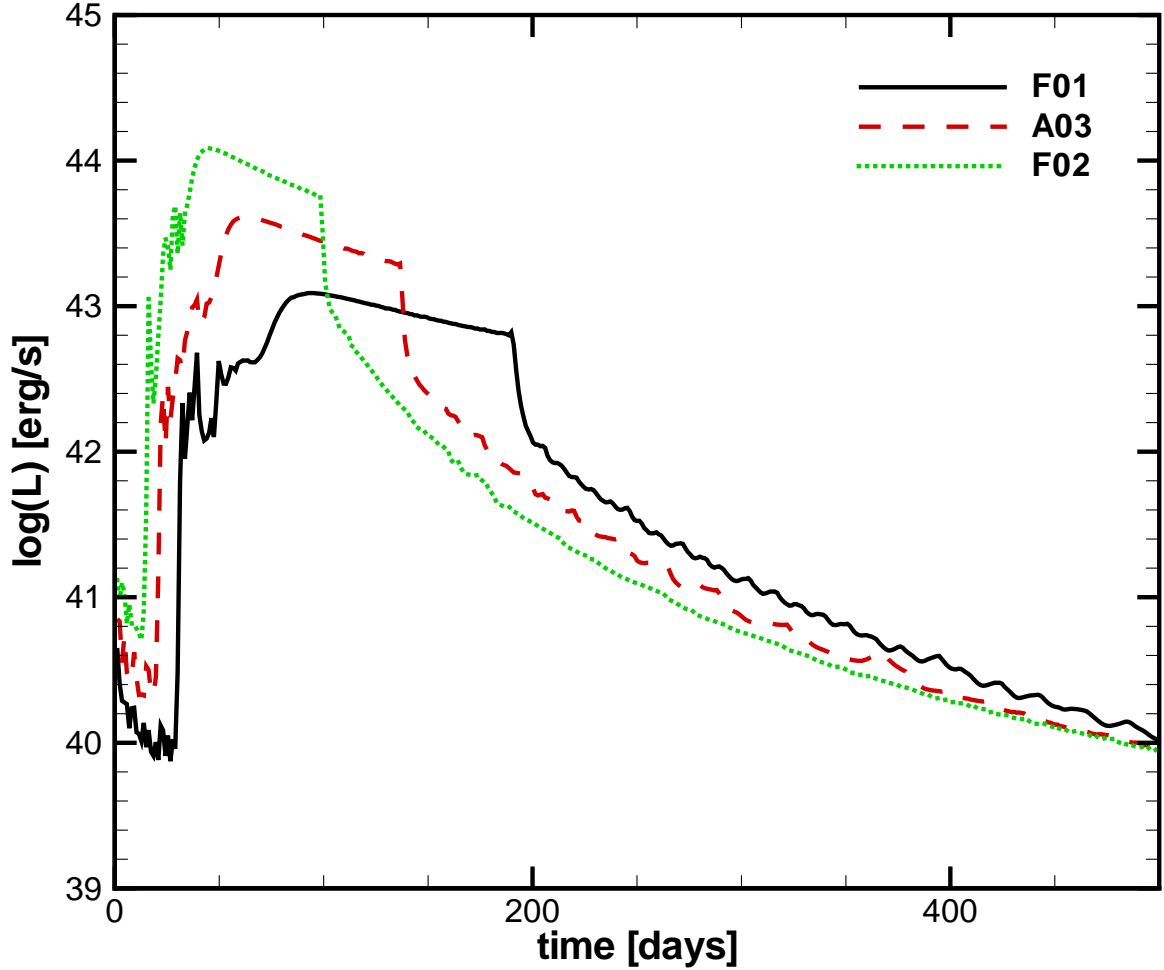}
  \caption{The influence of SN kinetic energy on the lightcurve. The
    higher the total energy, the higher the peak in the lightcurve and
    the shorter its duration, due to the increase in velocity.}
\label{fig:L_Esn}
\end{figure}

\subsection{Supernova Energy}

Altering the total energy of the initial SN explosion also can change
the apparent shape and luminosity observed in the light curve during
its collision with a circumstellar shell.  Indeed, in the case of
SN~2006gy, \citet{S09b} measure a total energy ($E_{rad}$ +
$_{kinetic}$) of at least 5$\times$10$^{51}$ erg.
Figure~\ref{fig:L_Esn} shows the lightcurves resulting from
simulations F01, A03, and F02, where three different SNe of the same
mass but kinetic energy of 0.5, 1, and 2 $\times$ 10$^{51}$ erg,
respectively, all collide with the identical circumstellar shell of
10~$M_{\odot}$ expanding at the same speed of 200 $\kms$.

In the discussion above, we found that higher SN ejecta speeds and
lower SN masses (a result of assuming that they all have the same
explosion energy of 10$^{51}$ erg) was a key factor contributing to a
high peak luminosity.  The key ingredient of higher ejecta speeds can
also be achieved with more total energy in an explosion, so we
explored this as well.  As one might naturaly expect, more energetic
SNe lead to higher peak luminosities because they give up more of
their initial energy as their faster ejecta suffer a sharper
deceleration during the collision.  The light curve peak is also
narrower (shorter in duration) for the more energetic and faster SNe
because it takes less time to overrun the same shell.

The net effect of altering the SN energy is similar to that of
changing the SN mass (but keeping the same energy), mainly because of
the strong influence of the SN ejecta speed (i.e. compare Fig 14 to
the middle panel of Fig 13).  Comparing F01, A03, and F02 in Figure 15
and Table 1, we see that SNe with higher explosion energy had higher
peak luminosities, but also more total radiated energy and higher
efficiency in converting shock energy into radiation, due to their
higher speeds as discussed above.  As we will see below, a major
difference between these three SNe of different initial kinetic energy
is seen in their final blast-wave speed after the shock overruns and
exits the circumstellar shell, providing a potentially useful
observational diagnostic (see \S 6).

\begin{figure}[t]
  \includegraphics[width=\columnwidth]{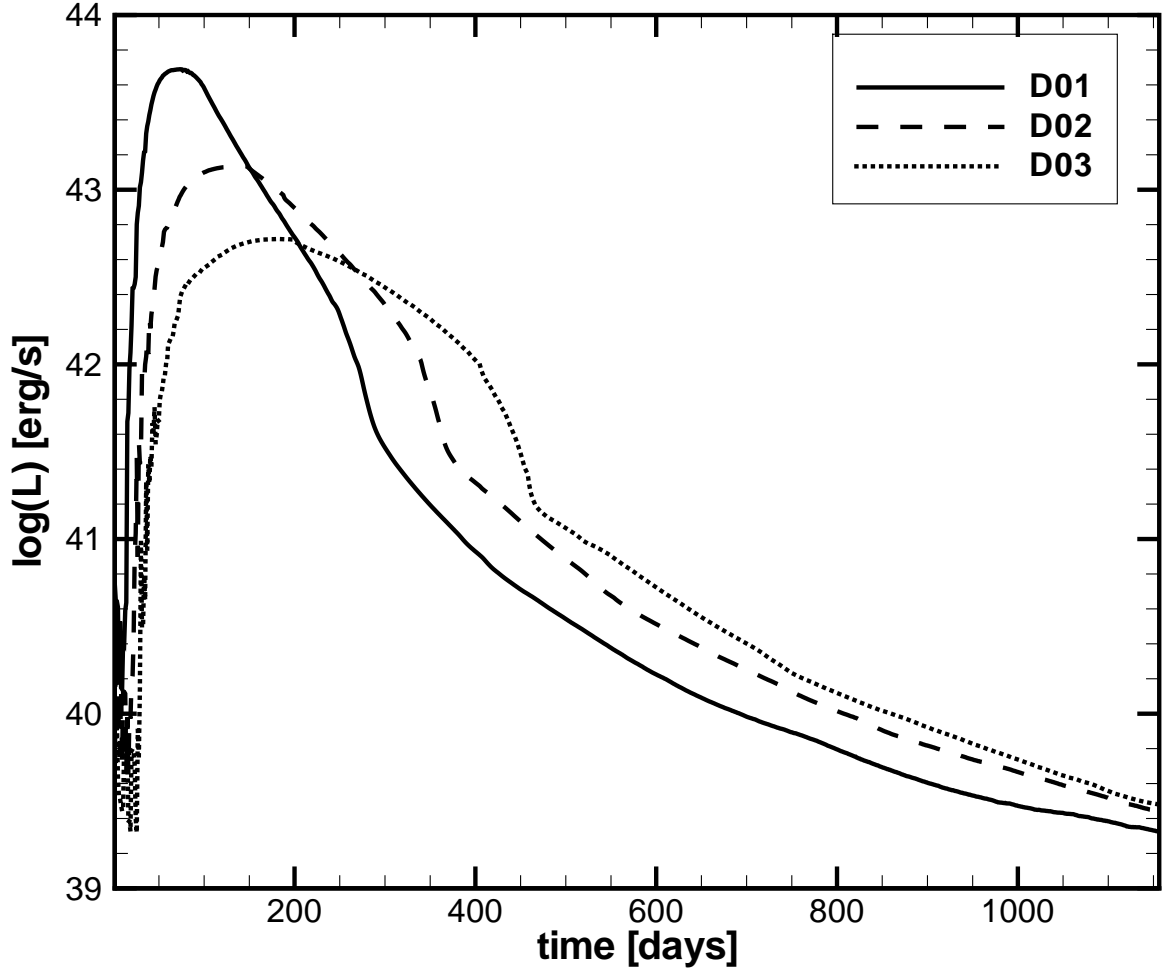}
  \caption{Bolometric lightcurves for the collision between three
    different supernovae with $m=10$~(B01), 30~(A09) and
    60~(C01)$\mso$ and a bipolar circumstellar shell. The Peaks in the
    lightcurve show the highest luminosity for the lowest mass
    supernova just as in fig.~\ref{fig:L_v_Msn}. The peaks are much
    roundar than for the collisions between supernovae and spherical
    shells.}
\label{fig:L_10bipol}
\end{figure}

\subsection{Bipolar nebulae}

So far, all our light curves have resulted from the collision between
a spherical SN and a spherical circumstellar nebula.  In contrast,
Fig.~\ref{fig:L_10bipol} shows the bolometric light curve produced by
the collision between the three SNe of three different masses and a
10~$M_{\odot}$ bipolar nebula.  Fig.~\ref{fig:L_10bipol} is analogous
to Fig.~\ref{fig:L_v_Msn}, but with a range of speeds in a single
shell as a result of its bipolar geometry instead of a range of speeds
in three different spherical shells. As with the spherical nebulae,
the lower mass SNe tend to produce higher peak luminosity in the
bipolar case because of their higher SN ejecta speeds.  However,
unlike the collisions between SNe and spherical nebulae, the
luminosity peaks have smooth curves and more gradual slopes, somewhat
reminiscent of the lightcurve of SN~2006gy \citep{S07}.  In our
simulations, at least, this smoothness results from the bipolar shape
of the nebulae.  Rather than an instantaneous collision between the
SNe and a circumstellar shell, the interaction starts gradually, with
the collision beginning first at the equator and than eventually
spreading to the pole.  An analogous transition happens when the SN
breaks out of the shell.  Again, this happens first at the equator and
only much later in the polar region.  As a result, shocked gas regions
with radically different temperatures and densities can exist
simultaneously, as shown in Figs.~\ref{fig:D02_10} through
\ref{fig:D02_150}.  One might imagine that a smoother light curve may
also result from a smoother transition in density at the outer
boundary of the shell.

A side effect of this situation would be that the possible onset
    of X-ray emission would be more gradual and not coincide with the
    drop in total luminosity as the SN breaks out of the shell.  The
    X-ray curves are expected to be strong when the supernova collides
    with a wind rather than a shell, both due to higher shock velocity
    and lower optical depth.  If the shell is spherical this
    transition happens everywhere at the same time.  In the case of
    the bipolar shell, the supernova breaks out at the equator long
    before it can break out at the pole.  Therefore, part of the shock
    may already generate observable high energy photons, while another
    part is still plowing slowly through the shell and emitting at
    much lower temperature with all high energy emission being
    absorbed.  Again, this effect may be relevant to the well-studied
    case of SN~2006gy, where the progenitor mass-loss rate inferred
    from the observed X-ray emission and H$\alpha$ luminosity is in
    severe disagreement with the mass-loss rate needed to power the
    continuum luminosity in a CSM interaction scenario \citep{S09b}.
    From fig.~\ref{fig:D02_10} through \ref{fig:D02_150}, one might
    understand this apparent contradiction if, for example, the X-rays
    are generated at the equator where the forward shock has already
    broken through the shell, whereas in the polar region the shock is
    still plowing through the dense massive shell and thereby powering
    the continuum luminosity.

\begin{figure}[t]
  \includegraphics[width=\columnwidth]{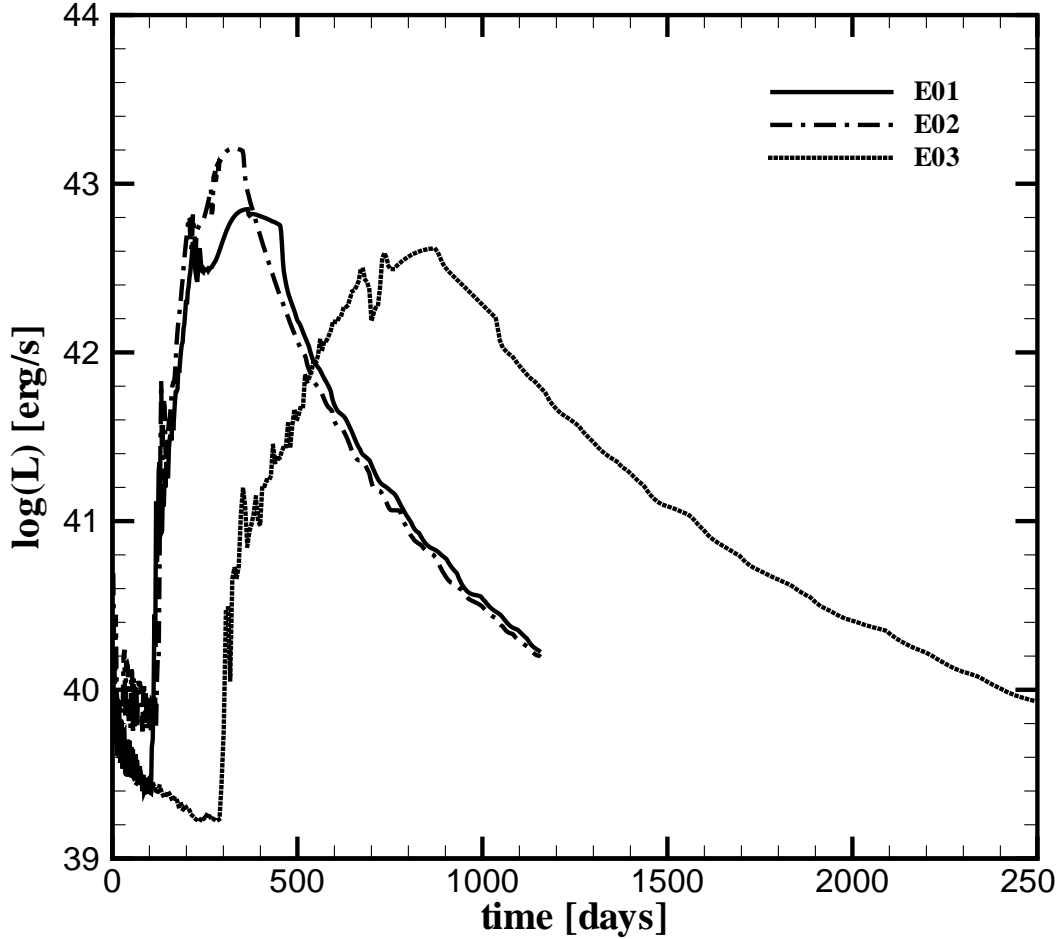}
  \caption{Bolometric lightcurves for the collision a 30$\mso$
    supernova and three 10$\mso$ shells, ejected at different times
    and velocities. The parameters of simulations E01 and E02 have
    been chosen in such a way that the inner boundary of the shell is
    at the same position, though moving with different velocities.  As
    a result, the lightcurves are nearly identical.  Simulation A03
    shows the lightcurve that results from collision with a shell that
    is much farther away from the star.  As a result the peak in the
    luminosity is much shallower.}
\label{fig:L_time}
\end{figure}

\begin{figure}[t]
  \includegraphics[width=\columnwidth]{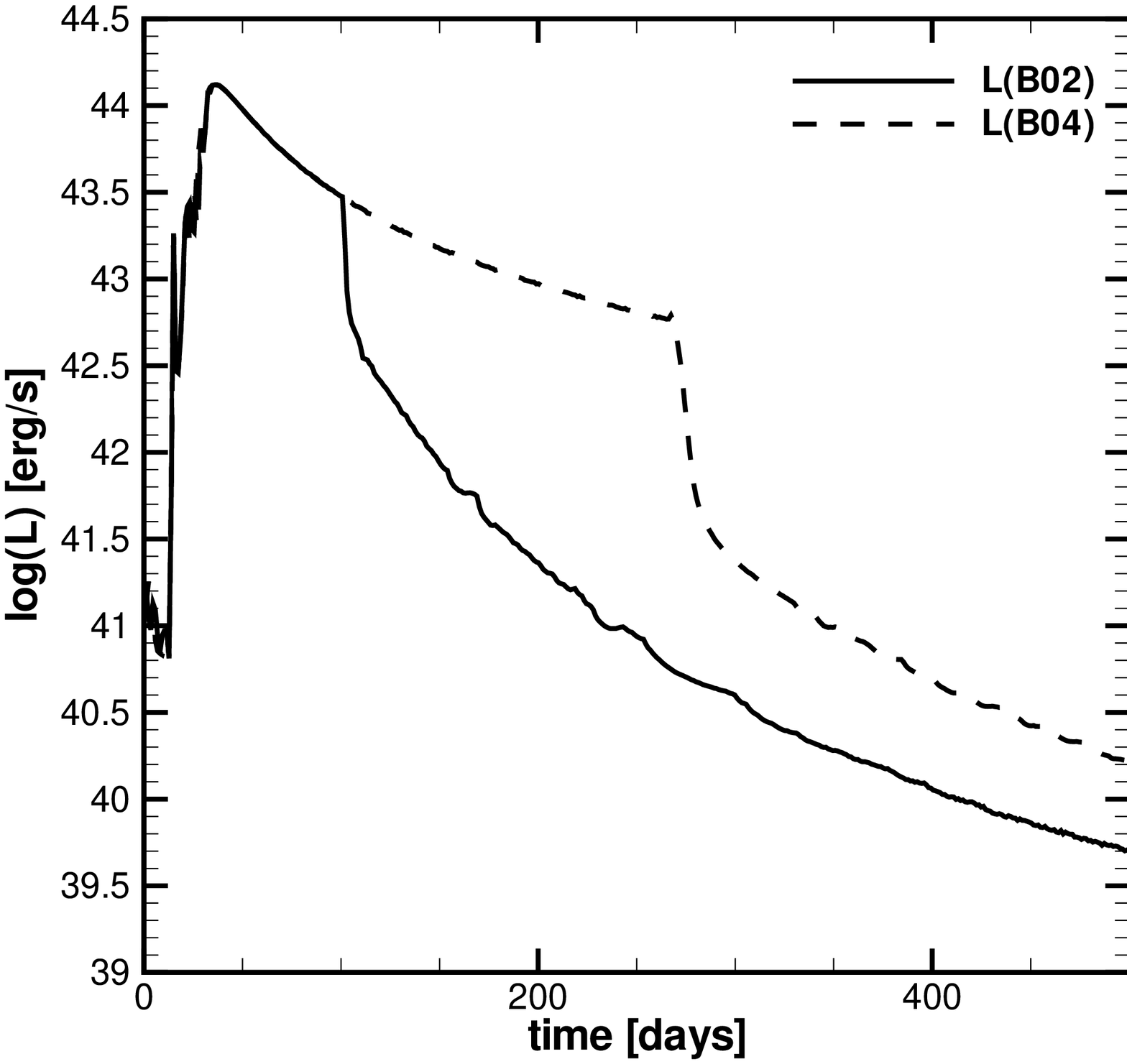}
  \caption{Total luminosity for simulations B02 and B04, demonstrating
    the effect of having a shell ejection that lasted over a longer
    period of time, making a thicker shell.  These simulations are
    identical at first, except that the CSM shell in B04 has a larger
    outer radius at the same density, and thus has a higher total
    shell mass and remains at high luminosity for a longer time as the
    blast wave plows through this additional material.  Since the
    cross section of the shell is larger for B04 the peak in
    luminosity lasts much longer.}
\label{fig:V_time2}
\end{figure}

\subsection{Time frames}

If the time interval between shell ejection and the SN changes, this
too will influence the shape of the light curve.  We investigate this
effect with simulations E01, E02, and E03, with the resulting light
curve shown in Fig.~\ref{fig:L_time}.  If there is a longer interval
of time between the precursor shell ejection and core collapse, there
are two effects.  First, a given shell can travel further away from
the star and will therefore have a lower density for the same shell
mass.  This will reduce the peak luminosity resulting from the shock
interaction.  The second effect of a larger time lag between shell
ejection and core collapse is that it delays the onset of the strong
CSM interaction phase.  This can potentially lead to a second
light-curve peak if one also considers the initial rise and fall of
photospheric emission of the underlying SN that we do not include
here.

For simulations E01 and E02, the shell velocities and ejection times
have been chosen so that the inner boundary of the circumstellar shell
is at the same position for both shells, though they are moving at
different speeds.  As a result, the lightcurves are quite similar in
onset and duration despite their different speeds.  Due to the
difference in shell velocity, the shell in E02 is denser than in E01,
resulting in a higher luminosity peak, which, however, quickly
disappears as the shock slows down.  Generally, the lower density of
these shells ($\rho\sim1/r^2$), results in lower luminosity peaks more
than 1 yr after core collapse, with edges that are less steep.  They
do not show the round peaks observed for bipolar shells
(\ref{fig:L_10bipol}).  This effect is seen most clearly in lightcurve
of simulation E03, which is extremely slow in its evolution, remaining
luminous for several years.

Another parameter is the outer boundary of the massive shell,
determined in our simulations by the duration of the shell ejection
episode and the speed of the shell.  In all simulations discussed so
far, the duration of the shell ejection phase was $\Delta t$=2 yr, and
we varied the outer radius of the shell by adjusting the speed of the
shell.  However, the {\it duration} of the shell ejection can vary as
well.  The 19th century eruption of $\eta$ Carinae, for example,
lasted about 20 yr \citep{dh97}, although the mass ejection seems to
have been concentrated in a shorter time interval \citep{S06}.  If the
shell ejection occurs at the same mass-loss rate but lasts longer,
then the shell will be thicker and have a larger total mass.  The main
effect of this is that the main peak of the light curve would last
longer at a comparable luminosity. This is indeed the case, as we show
in Fig~\ref{fig:V_time2}, which compares simulations B02 and B04.
This is different from the case mentioned above referring to the speed
of the shell. If a larger outer radius and longer duration to the
light curve result from a faster shell speed, then the density is
lower for the same mass and the resulting luminosity will be much
lower (compare simulations B01, B02, and B03).

\begin{figure}[t]
  \includegraphics[width=\columnwidth]{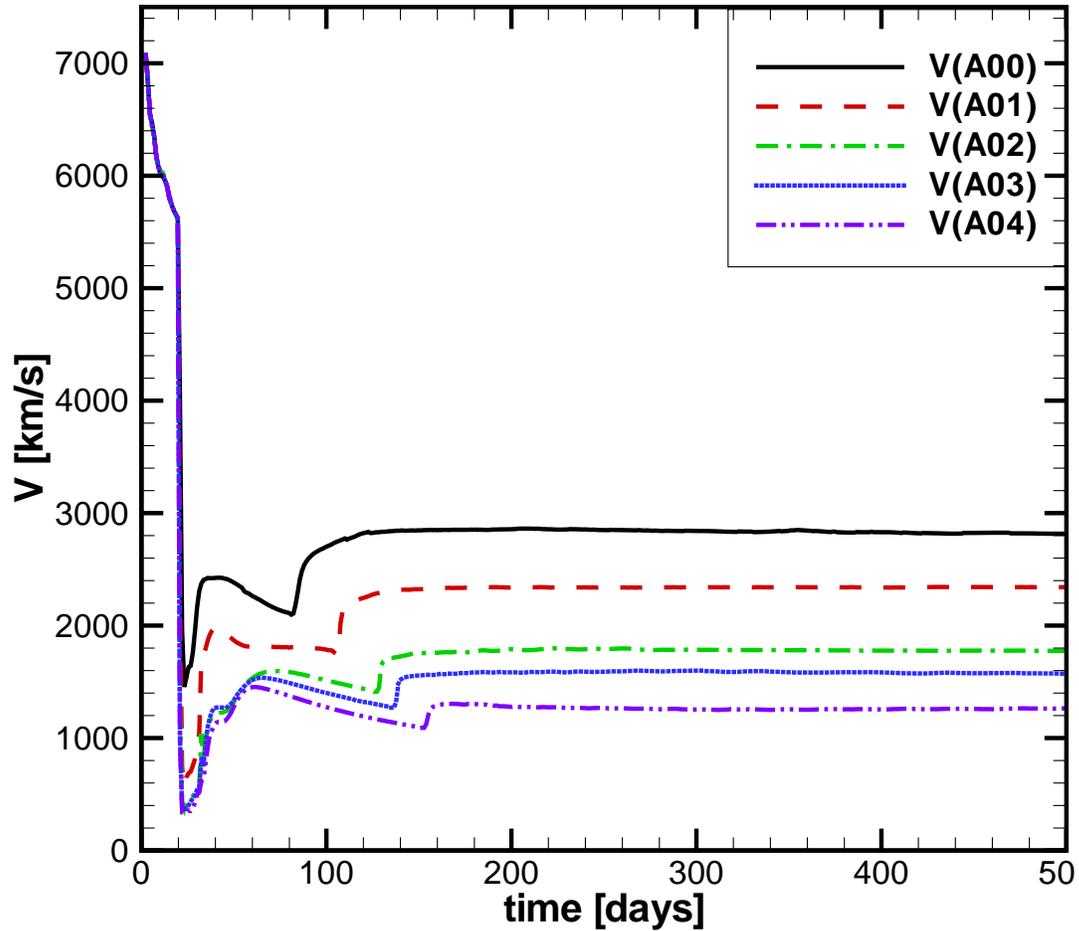}
  \caption{Shocked gas velocities for the same simulations as in
    fig.~\ref{fig:L_mshell}.  The velocity drops very abruptly when
    the supernova hits the circumstellar shell, then rises again as
    the supernova plows through the gas, then rises once more as it
    breaks out and makes the transition from a radiative shock back to
    an adaiabatic one. The final velocity depends clearly on the mass
    of the shell.}
\label{fig:V_mshell}
\end{figure}

\begin{figure*}
 \centering
\resizebox{\hsize}{!}{\includegraphics[angle=-90,width=0.95\textwidth]{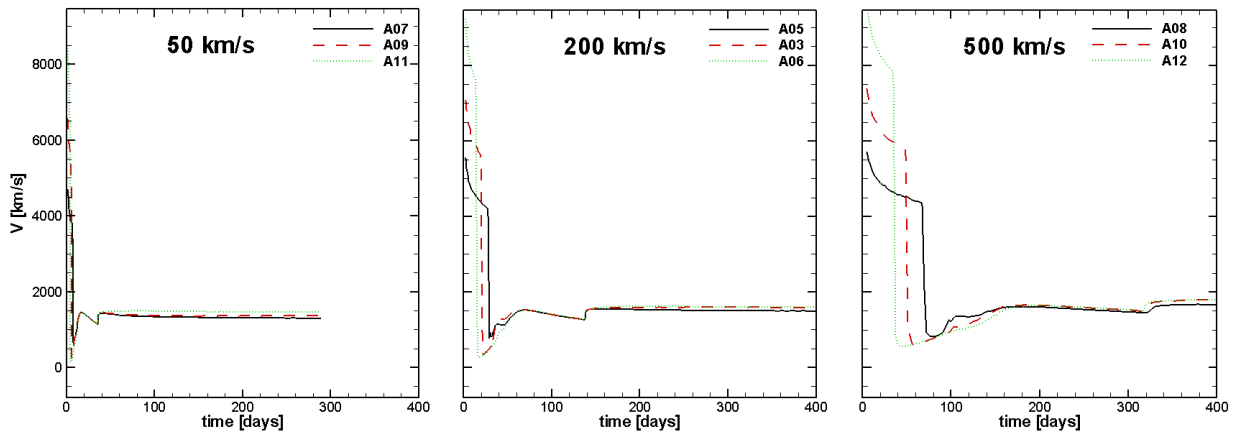}}      
\caption{Shocked gas velocities for the same simulations as in
  fig.~\ref{fig:L_v_mdot}.  The wind velocity changes the entire shape
  of the lightcurve, as it changes the location of the shell, whereas
  the wind mass loss rate is only important in the initial stages.}
 \label{fig:V_v_mdot}
\end{figure*}

\begin{figure*}
 \centering
\resizebox{\hsize}{!}{\includegraphics[angle=-90,width=0.95\textwidth]{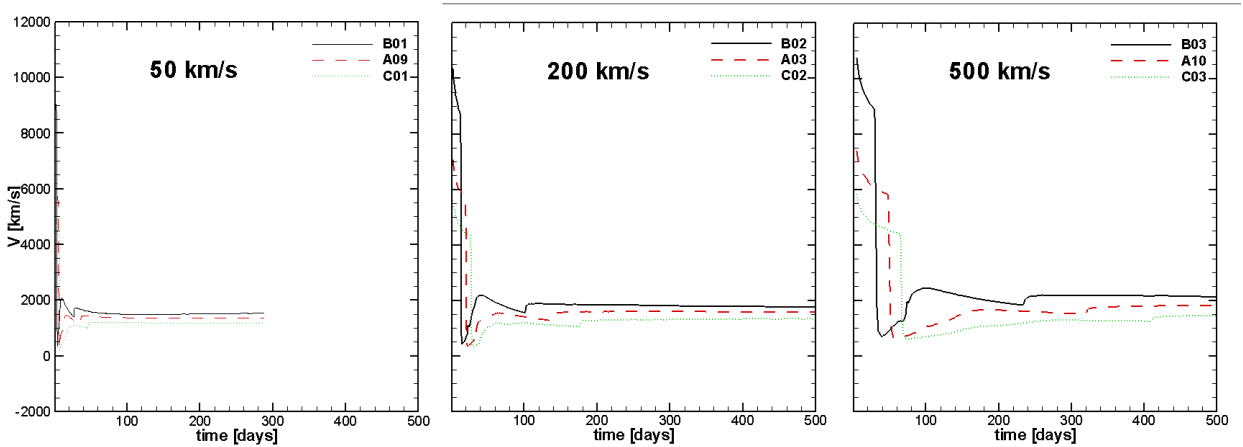}}      
\caption{Shocked gas velocities for the same simulations as in
  fig.~\ref{fig:L_v_Msn}.  The high mass supernovae, which have
  relatively low velocity slow down less than the low mass supernovae,
  but not enough to reverse the shock velocities.}
 \label{fig:V_v_Msn}
\end{figure*} 

\begin{figure}[t]
  \includegraphics[width=\columnwidth]{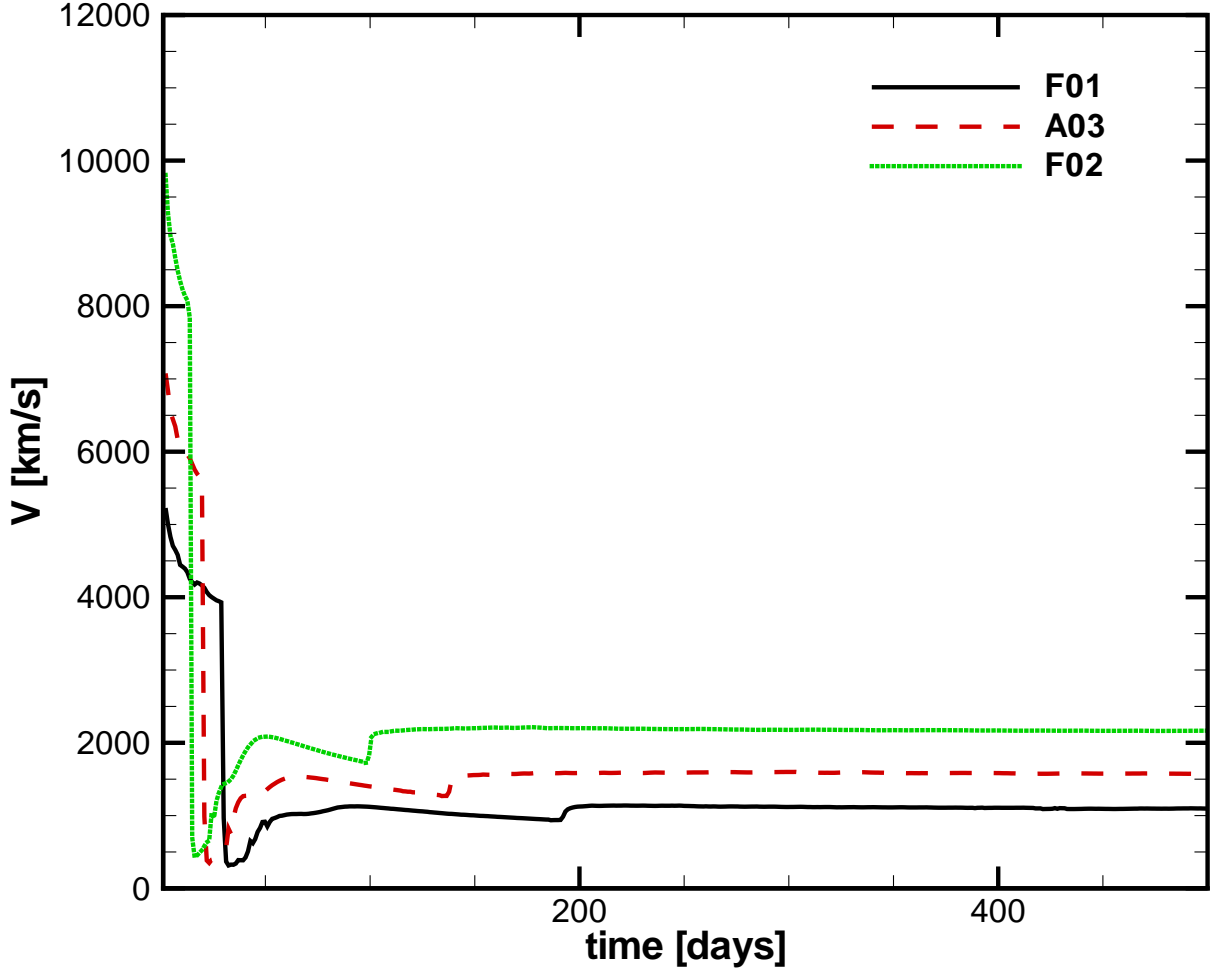}
  \caption{Reverse shock velocities for the same simulations as in
    fig.~\ref{fig:L_Esn}, illustrating the effect of SN energy on
    $v(t)$. The more rapid deceleration of F02 occurs simply because
    the fastest ejecta reach the inner boundary of the dense CSM shell
    sooner.  The higher energy SN has higher speeds at later phases
    during the collision.}
\label{fig:V_Esn}
\end{figure}

\begin{figure}[t]
  \includegraphics[width=\columnwidth]{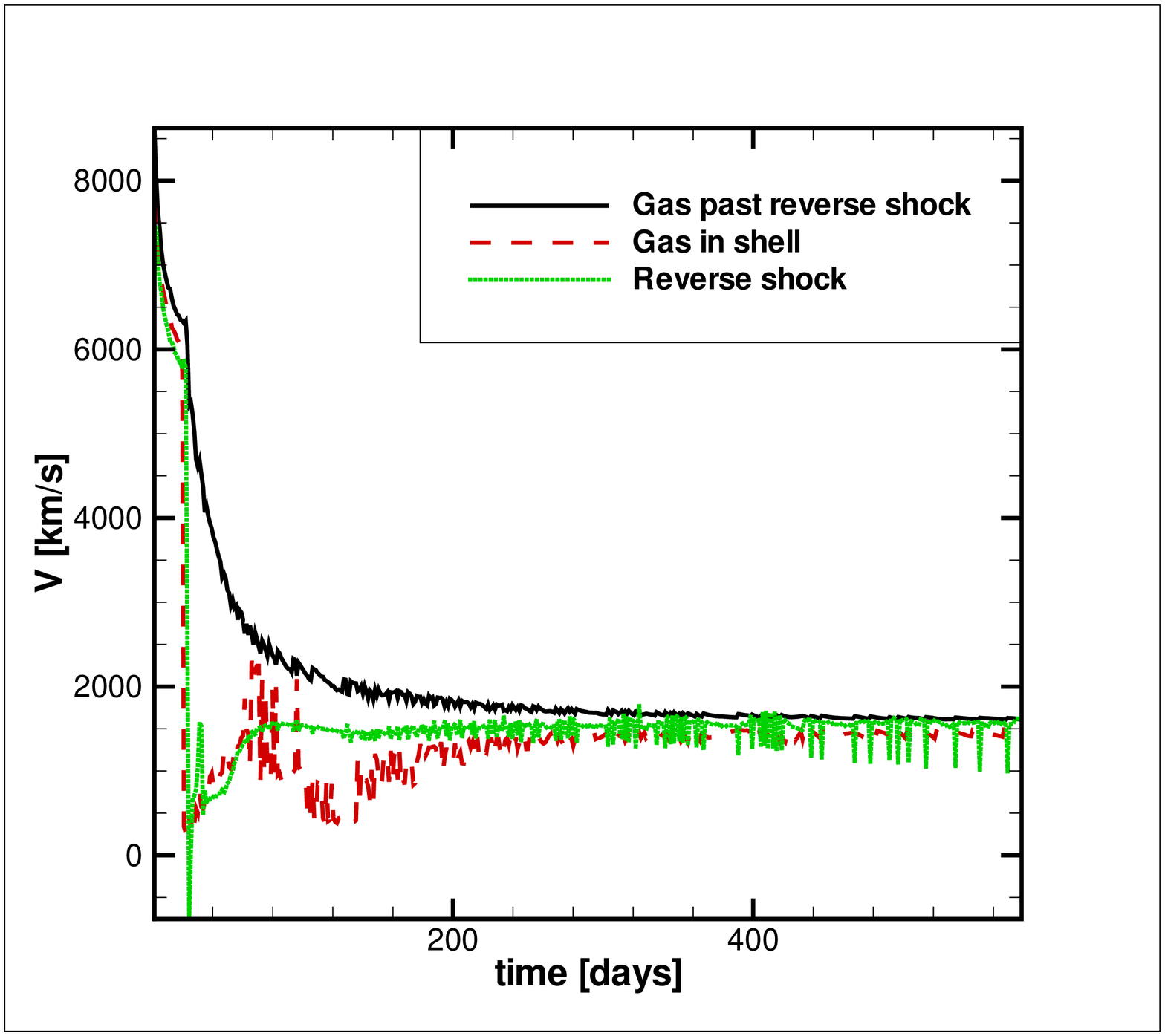}
  \caption{velocity curves for the reverse shock, shocked gas flow
    past the reverse shock and highest density shocked gas  (the shell) for
    simulation A03.}
\label{fig:V_A03}
\end{figure}

\section{Supernova shell velocity}

An important observational parameter for SNe~IIn, in addition to their
radiative luminosity and total radiated energy measured from light
curves, is the expansion speeds measured from line widths in
spectra. In most SNe, the ejecta expansion speeds are inferred from P
Cygni absorption features in the photospheric spectra, and this can be
done in SNe IIn if the underlying photosphere can be seen
\citep{CD94,T93,S02,S09a}.  Often, however, the underlying SN
photosphere is masked by the bright and possibly opaque emission from
the dense shell of post-shock gas that powers the excess luminosity in
SNe~IIn (e.g., \citealt{Cetal04,S08a}).  Fortunately, the dense shell
of shocked gas that piles up at the contact discontinuity in the
SN-CSM collision can be seen in the intermediate-width wings of the
narrow H$\alpha$ emission lines, for example, and typically has a
speed of a few 10$^3$~$\kms$ \citep{CD94,T93, Cetal04, F02, P07,
  S07,S08a,S08b,S09a,S09b}.

In order to estimate how the presence of a circumstellar shell
influences the velocity of the post-shock gas seen in H$\alpha$
emission, we plot the velocity of shocked gas as a function of time:
specifically, we plot the mass-averaged radial velocity of the gas
between the forward shock ($R2$) and the reverse shock ($R1$)
\begin{equation}
v_{\rm av}~=~\frac{\int_{R1}^{R2} \int_0^\pi  r^2 \sin(\theta)\rho v_r dr d\theta}
{\int_{R1}^{R2} \int_0^\pi r^2 \sin(\theta)\rho dr d\theta}.
\end{equation}
We choose this method to quantify the shocked gas velocity because it
gives a good result both in the adiabatic and radiative shock regime.
Mass averaging rather than volume averaging is more realistic since
the luminosity is highly density dependent, so high density areas
would dominate the emission.

The behavior of the shocked gas velocity during the SN-shell collision
generally proceeds as follows: Initially, the velocity drops
exponentially, because the blast wave declerates while sweeping up the
surrounding wind.  When the forward shock hits the circumstellar shell
it practically halts and the reverse shock velocity drops abruptly as
the gas between the two shocks is compressed.  After the initial
collision the shocked gas velocity increases again as the forward
shock recovers.  However, the velocity is now much lower since the
interaction has become radiative, so much of the available energy has
already been lost.  Also, the forward shock is now moving through a
much denser medium.  Although the forward shock accelerates again as
it breaks out of the shell and runs down a steep density gradient, it
never recovers its original velocity, as is true for the reverse
shock.  The forward shock interaction does become nearly adiabatic
again (see \S~\ref{sec-sn-csm}), so the velocity remains higher than
during the collision with the shell.  By this time the shell has
gained substantial mass through sweeping up the surrounding medium.
Therefore its forward momentum is high, and the velocity remains
nearly constant for a long period of time because the outer wind has
insufficient mass to decelerate it.

The abrupt loss of forward velocity in the reverse shock shell is, in
principle, a robust characteristic of SN-shell interaction.  Whether
or not it is actually observable, however, is unclear.  If the density
of the pre-shock CSM is high, as it needs to be in the case of the
more luminous SNe IIn, then one might expect the pre-shock gas to be
very optically thick \citep{SM07,S09b} and the emitting surface may be
well outside the shock.  In that case, the observed H$\alpha$ line
profile would be dominated by the narrow component from photoionized
pre-shock gas (typically a few 10$^2$~$\kms$) and broad electron
scattering wings (e.g., \citealt{C01,D08}) out to a few 10$^3$~$\kms$.
This is indeed thought to be the case for SN~2006gy, as discussed in
detail by \citet{S09b}.

The {\it final} shocked gas velocity, on the other hand, should be
easily observable in all cases because of lower optical depths at
larger radii and at late times, and may therefore provide an
unambiguous constraint on the CSM mass and SN energy.  In
Figure~\ref{fig:V_mshell}, which shows the velocity for the same
simulations as fig~\ref{fig:L_mshell}, we can see that the final
velocity does in fact depend strongly on the mass of the circumstellar
shell.  If the shell mass is relatively high, the velocity decreases
by a larger amount as momentum is conserved.

In fig.~\ref{fig:V_v_mdot}, we show the shocked gas velocity for the
same simulations as Figure~\ref{fig:L_v_mdot}.  This demonstrates the
effect of the wind velocity and mass-loss rate on the shocked gas
velocity.  As all SNe have the same total energy in these simulations,
the higher mass SNe start out with lower velocity.  As can be seen,
the mass-loss rate only matters in the initial stage, before the
collision between the SN and the circumstellar shell. The wind
velocity does make a significant difference as it determines the
location of the shell relative to the star and therefore the timetable
of the interactions, but it does not strongly influence the final
speed of the shocked shell.

In fig.~\ref{fig:V_v_Msn}, which shows the reverse shock velocity for
the same simulations as Figure~\ref{fig:L_v_Msn}, we demonstrate the
effect of the mass of the SN ejecta on the reverse shock velocity.  As
all SNe have the same total energy in these simulations, the higher
mass SNe start out with lower velocity.  As the ejecta collide with
the 10$\mso$ shell, the lower mass SNe slow down more, since they have
less momentum.  Still, after the collision, the lowest mass SN is
still moving with the highest velocity and is the first to break out
of the circumstellar shell.

Finally, fig.~\ref{fig:V_Esn} shows the reverse shock velocities for
SNe with the same mass, but different total energy.  The high energy
SNe start out with higher velocities, but also lose more energy in the
collision.  Since the collision takes place at an earlier stage, they
slow down more, so the final difference in velocities is much less
than initially.  However, the total SN kinetic energy is perhaps the
most influential factor in determining the final shell speed.

Because of the specific nature of the collision, their are several
different features that can determine the observed shock velocity of
the SN.  To illustrate this effect, we show three alternative velocity
curves in Fig.~\ref{fig:V_A03}, all based on simulation A03.  A) The
velocity of the reverse shock, B) The velocity of the gas that has
passed through the reverse shock, and C) The velocity of the shocked
gas that has the highest density.  Initially, all three curves move
together.  The gas flow of shocked gas is slightly faster than the
reverse shock itself, since this gas is actually moving through the
shock.  Before the collision with the circumstellar shell the highest
density of shocked gas is at the reverse shock, since this is where
the SN ejecta piles up.  This changes once the supernova collides with
the shell.  The reverse shock recoils from the collision, stopping
completely or even reversing, depending on the density of the shell.
The gas velocity inside the shock decreases as well, but not as much,
since this is governed by the shock conditions.  The shock changes
from adiabatic to isothermal, restricting the velocity jump over the
shock.  As the SN then ploughs through the shell, the location of the
highest density feature changes.  It is no longer at the reverse
shock, but rather at the forward shock, where gas from the shell is
being swept up.  Therefore, the flow speed of the high density feature
actually becomes lower than the velocity of the reverse shock, since
we are now sampling gas that is still in the process of being
accelerated.  Once the blast wave breaks out of the shell, the
original situation is recreated, as once more the highest density
occurs at the reverse shock and the shock conditions change back from
isothermal to adiabatic.

Examining Figures~\ref{fig:V_mshell} %%20
through \ref{fig:V_A03}, %% 24
one can see that velocities measured in spectra obtained at early
times can be powerful diagnostics of the rapid changes occurring
during the initial shell collision, while later spectra that provide
estimates of the final coasting velocity of the CDS {\it are key
  diagnostics of the energy and momentum budget of the explosion}.  A
potential complication for the early-time velocities, especially with
more luminous SNe~IIn, may arise of the inner CSM is very optically
thick.  If the CSM outside the shock is highly opaque, then a
radiative precursor may cause the photosphere to reside outside the
shock \citep{S09b}, in which case the observed velocities are not
indicative of the true expansion speed.  In any case, combinations of
photometry and spectra at early times while the SN is still on the
rise to peak are quite valuable in breaking the degeneracy of various
models.

\section{Discussion}

\subsection{The Influence of SN and Shell Properties}

In the previous sections we have shown how circumstellar shells can
influence the evolution of both the observed SN lightcurve and the
observed velocity.  The presence of a substantially dense
circumstellar shell always causes an increase in the radiative
luminosity, lasting until the blast wave breaks through the shell.
The height of this luminosity peak depends primarily on the density of
the shell (and so, also on its total mass and speed), in the sense
that denser shells invariably lead to higher luminosities for the same
underlying SNe.  The duration of the luminosity peak is a direct
consequence of the time it takes the blast wave to propagate though
the shell, so it depends on the total mass of the shell, its expansion
speed, and its inner and outer radii (i.e. the \emph{duration} of the
pre-SN ejection episode).  A relatively more massive shell produces a
slower blast wave, increasing the duration of the light curve peak and
causing a higher luminosity.  A faster expansion speed for the shell
will also stretch the duration of the light curve peak by increasing
its outer radius, but will make it less luminous for the same mass.

In our simulations, typical luminosity peaks for spherical shells tend
to have a flat plateau, which is either horizontal, or angled downward
as the shock velocity decreases over time.  The beginning and end of
the light curve peaks are clearly defined with sharp edges, but this
is just a result of our simplifying assumption that the shell has
sharp inner and outer boundaries; real shells may have more
complicated density profiles.  These characteristics tend to disappear
if the nebula is bipolar in shape, because different latitudes in the
bipolar shell are hit by the blast wave at different times, and so the
light curve shape is smoother.

Whereas the total luminosity and the visual luminosity peak when the
blast wave collides with the circumstellar shell, the temperature of
the emitting gas decreases as the shock slows.  A gradual decline in
the characteristic temperature inferred from the continuum slope in
visual-wavelength spectra or multi-band photometry has been seen in
several well-studied examples of very luminous SNe~IIn, such as
SN~2005gj, SN~2006gy, and SN~2006tf \citep{P07,S08a,S09b}.  The X-ray
drop may not be observed if initial phases are optically thick and
X-rays are fully absorbed and reprocessed.  Trapping at high optical
depths is an effect that we have not included directly in our
simulations; we consider it likely, therefore, that the visual
radiation will trace the bolometric luminosity at early times, as we
discussed earlier.  This is why we have shown the bolometric
luminosity light curve in our plots.  Luminosity at high energies
increases again once the SN breaks out of the shell and interacts with
the (relatively) low-density wind outside the shell.  This change in
shock temperature is less clearly defined if the shell is bipolar,
because both high-velocity and low-velocity interactions can occur
simultaneously in different parts of the shell.  A clumpy CSM may
produce a similar effect.

The observed velocity evolution of the dense post-shock H shell
depends strongly on the CSM density and SN energy in our simulations.
This velocity decreases steeply in the earliest phase of the expansion
when the blast wave sweeps through the wind inside the dense CSM
shell, and then it takes another drop when the shock hits the
circumstellar shell.  However, these velocities in the earliest phases
may be difficult to observe because of high optical depth effects that
are not taken into account in our simulations, as noted above for the
early light curve shape.  The characteristic velocity observed {\it
  after} the SN/shell collision ends depends on the shell mass, SN
mass, and the total explosion energy, and is typically 1--3 $\times$
10$^{3}$ $\kms$ in our simulations.  This is comparable to the
observed linewidths in luminous SNe~IIn like SN~2006tf \citep{S08a} 
or SN~2005gj \citep{P07}.  The faster speeds of
$\sim$4,000 km s$^{-1}$ in SN~2006gy \citep{S09b} imply a
higher energy explosion and a relatively high-mass SN.  
Indeed, \citet{S09b} estimated an explosion energy of at least
5$\times$10$^{51}$ ergs for SN~2006gy.

The mass and initial speed of SN ejecta (and hence, the total
explosion energy) also influence the evolution of the velocity.  SNe
with higher ejecta mass have higher inertia and are decelerated less,
but they also have slower initial expansion speeds for explosions
assumed to have the same total kinetic energy, and so they can end up
with slower final expansion speeds.  A more energetic and relatively
more massive SN explosion will emerge from the shell collision episode
with a faster final shock speed.  Since there is some degeneracy in
any one type of observed property, spectral observations of the
pre-shock CSM speed, the post-shock shell, and the SN ejecta speeds
(if they can be seen) are valuable to combine with estimates of the
luminosity from photometry to derive the physical properties of the
CSM interaction.

\begin{figure}
%%
%% plot efficiency.eps
%%
\includegraphics[width=\columnwidth]{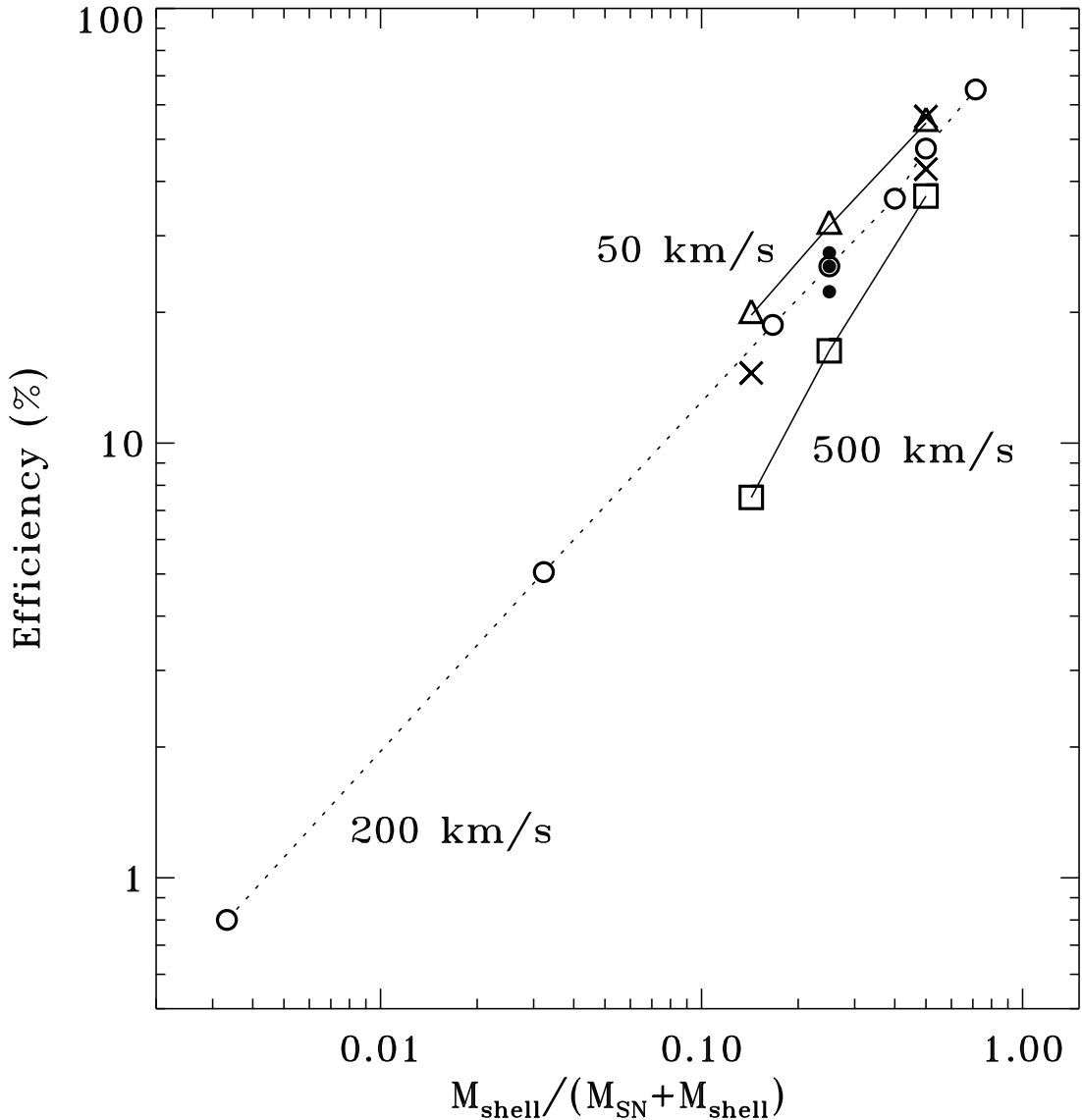}
\caption{The total efficiency (\%) in converting shock kinetic energy
  into radiated energy from Table 1 for several representative
  simulations, plotted as a function of the ratio of CSM shell mass to
  the total mass involved (SN ejecta + shell).  The unfilled circles
  (and dotted line) represent our baseline simulations with SN ejecta
  with 10$^{51}$ erg running into 200 km/s shells of various masses
  (``A'' models, plus B02 and B04).  The unfilled triangles and
  squares are similar but for CSM speeds of 50 and 500 $\kms$,
  respectively.  The filled circles are models F01, A03, and F02,
  showing the effect of different explosion energy for the same shell
  paramters.  The X's show models G01, H01, and C01 (all with
  $V_{exp}$ = 200 km s$^{-1}$), special cases that have extended CSM
  mass or lower-mass SN ejecta.}
\label{fig:efficiency}  
\end{figure}

\subsection{The Shock Conversion Efficiency}

Since some very luminous SNe~IIn have measured values for their total
radiated energy approaching or even exceeding the canonical SN
explosion kinetic energy of 10$^{51}$ erg, the efficiency at which
they convert some fraction of their initial kinetic energy into
post-shock thermal energy and then radiation is key.  In the
shell-shocked model \citep{SM07}, high efficienies are allowable
because of the large radius at which shock energy is thermalized,
allowing the SN to radiate before it expands and loses that thermal
energy adiabatically.  The second to last column of Table 1 lists the
efficiency of this conversion as the ratio of the total energy lost
via radiation in each simulation to the initial explosion kinetic
energy of the SN, or $dE/E_{SN}$.  In Figure~\ref{fig:efficiency} we
plot this efficiency as a function of another ratio, which is the CSM
shell mass compared to the total mass in both the SN ejecta and CSM.
We show the results for several simulations to demontrate various
trends.

The basic result is that efficient conversion of SN kinetic energy
into radiation via CSM interaction requires a CSM mass that is
comparable to or larger than the mass of the SN ejecta.  The primary
criterion for luminous SNe~IIn that result from core-collapse SNe is
therefore the presence of several $\mso$ of circumstellar gas which
must have been ejected very shortly before the SN.  Explosions of very
massive stars can have CSM interaction that is not very luminous if
the CSM mass is small compared to the SN ejecta mass (as long as the
SN ejecta are slow and heavy for a standard energy).  A very effective
way to convert a larger fraction of the total initial energy (more
than half) into radiation is to have a more extended CSM shell at the
same density, as in simulation B04, tracing mass loss for a longer
time prior to the SN explosion.  Of course, the longer a simulation
runs into CSM material, the more kinetic energy can be converted into
light --- if one waits for $\sim$100 yr or more, an extended SN
remnant can tap a significant fraction of the total energy.  Our aim
here, however, is to study objects that do this very quickly in
$\sim$1 yr and thereby produce high luminosities during the initial
light curve peak.

The pre-shock CSM speed also has some minor effect on the efficiency,
in the sense that slower CSM speeds lead to denser environments that
trap more of the available kinetic energy because of their denser
post-shock gas, and consequently, more efficient cooling.  Also mildly
influential is the speed of the SN ejecta, or equivalently, the SN
explosion kinetic energy.  More energetic explosions are more
efficient in converting their available energy reservoir to radiation
due to the higher velocity drop at the reverse shock.  Thus, mild
increases in explosion energy offer an alternative to exceedingly
massive CSM shells in order to produce very luminous SNe~IIn.  Again,
however, the CSM must be extended and massive in order to maintain
that high luminosity for an extended time.

\begin{figure}
%%
%% plot lightcure06tf.eps
%%
  \includegraphics[width=\columnwidth]{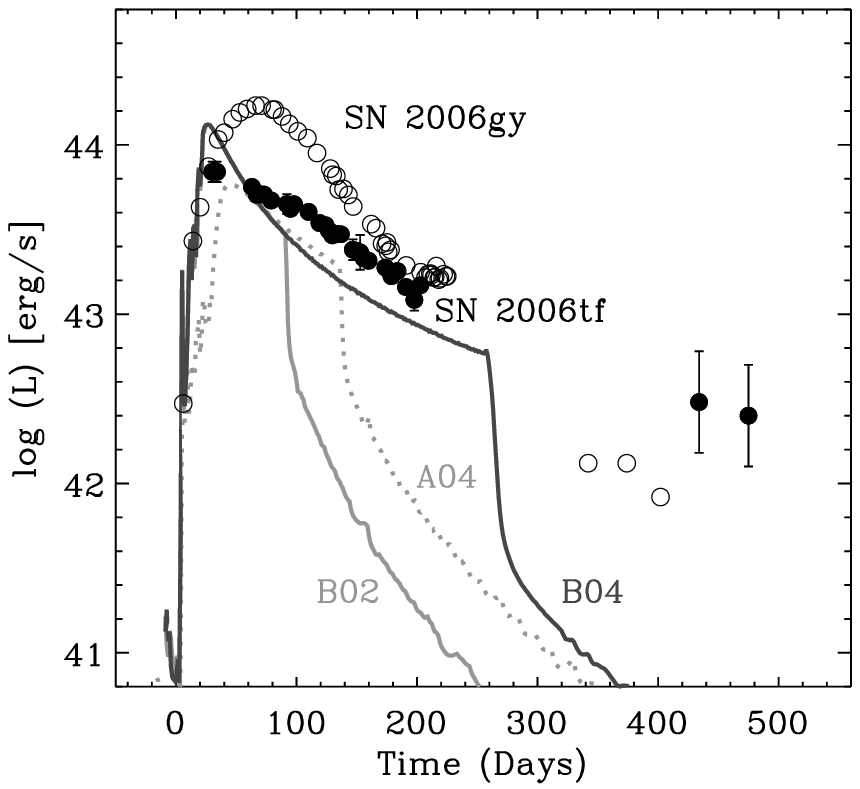}
  \caption{This plot compares a few selected models to light curves of
    the two most luminous SNe~IIn known.  SN 2006gy (data from
    \citet{S07,S08c}) is shown with unfilled circles, and SN 2006tf
    \citep{S08a} is shows with filled circles.  The models shows are
    A04 (30 $M_{\odot}$ SN and 20 $M_{\odot}$ shell; dotted grey), B02
    (10 $M_{\odot}$ SN and 10 $M_{\odot}$ shell; solid grey), and B04
    (10 $M_{\odot}$ SN and 25 $M_{\odot}$ shell; dark grey).  Model
    B04 is the same as B02 except that the shell has a larger outer
    radius (it is thicker at the same density) and therefore has a
    larger total mass.  While these models may account for the peak
    luminosities of SN~2006tf, they fall short of the peak luminosity
    for SN~2006gy and they fade too quickly for both.  It is likely
    that simulations with more extended and more massive CSM shells or
    more energetic SNe need to be explored in these two particuar
    cases.  The very late time data at around 400 d for SN~2006gy may
    have some contribution from a light echo (e.g., \citet{S08c}), but
    the late-time luminosity after 1 yr for SN~2006tf is dominated by
    strong ongoing CSM interaction because strong H$\alpha$ emission
    is seen in the late-time spectrum \citep{S08a}.  For a typical
    Type IIp lightcurve, see Figs.\ \ref{fig:L_O0123} and
    \ref{fig:L_mshell}.}
\label{fig:compare}  
\end{figure}

\subsection{Comparisons to observations of luminous SNe~IIn}

The central motivation for this study was to explore shock interaction
with dense pre-SN CSM shells as a possible engine for the visual light
from the emerging class of extremely luminous SNe, and to ask whether
observed light curves can be compared to expectations of hydrodynamic
simulations in order to constrain the physical properties of those
shells and the underlying SNe.  Below we briefly mention two recent
well-observed examples that have been our primary objects for
comparison: (1) SN~2006gy \citep{S07,O07} was the first of these
super-luminous SNe that raised many questions about our understanding
of the power sources for these objects and about massive star
evolution, and (2) SN~2006tf \citep{S08a} which was nearly as
luminous.  Both have optical spectra of Type~IIn suggesting the
presence of dense CSM, although SN~2006tf appears to fit the canonical
picture of CSM interaction as the power source with fewer
complications \citep{S08a}.  Both appear to have high optical depths
at early phases.  The energy sources for these two SNe are of
particular interest because, unlike more common SNe IIn at lower
luminosity, their total radiated energy severely taxes the total SN
energy budget: the energy radiated in visual light was
0.7$\times$10$^{51}$ ergs for SN~2006tf \citep{S08a} and
1.6$\times$10$^{51}$ ergs for SN~2006gy \citep{S07}.  Making a
bolometric correction based on the observed temperature yields
$E_{rad} \simeq 2.5 \times 10^{51}$ erg for SN~2006gy, and including
the kinetic energy remaining in the shell pushes the total initial
explosion energy to at least 5$\times$10$^{51}$ erg \citep{S09b}.  For
SN~2006gy, a pair instability SN or diffusion from an opaque shocked
shell have also been suggested as possibilities for powering the
observed luminosity \citep{SM07,S07,O07,WBH07}, although detailed
analysis of its spectral evolution favors the opaque shocked shell
model \citep{S09b}.

Two other SNe, SN 2005ap \citep{qetal07} and SN 2008es
\citep{metal2009,getal2009}, have also been discovered recently to be
among the most luminous SNe known.  In fact, their peak luminosities
are somewhat higher than SN~2006gy, although they faded more quickly.
We do not consider these for direct comparison with the same type of
model discussed here because their spectra are not of Type~IIn, but
rather, they had normal broad-lines in their spectra indicating a
photosphere receding through fast ejecta.  The lack of narrow emission
lines makes it likely that their radiation is produced primrily by
diffusion from an opaque shocked envelope, as in the model of
\citet{SM07}, but with a smaller envelope mass than for SN~2006gy.
The parameters in some of our simulations with slower (and therefore
more dense) CSM and higher conversion efficiencies, such as B01, might
be appropriate for these objects if diffusion were properly accounted
for.  Alternatively, it has recently been suggested that these SNe may
powered by the birth of magnetars \citep{KB09,W09}, energizing the
opaque SN ejecta from within.

We did not tune models specifically to fit the observed light curves
of SNe~2006gy and 2006tf, but we did explore a range of parameters for
combinations of SNe and CSM shells comparable to relevant parameters
estimated from observations \citep{O07,S07,S08a,S09b,WBH07}.  Among
our models, some of the highest peak luminosities were attained with
relatively low mass (and therefore fast) SNe running into slower CSM
shells, such as B01 and B02 (10 $M_{\odot}$ SNe), or A07, A09, and A11
(slower 50 km s$^{-1}$shells).  Although these models achieved very
high peak luminosities comparable to those of the most luminous
observed SNe~IIn, they faded too quickly, and so they fall far short
of achieving the duration and total radiated light output of events
like SNe~2006gy and 2006tf.  Lower mass-SNe run out of momentum too
quickly, or over-run the compact CSM shells too quickly.  These low-
and moderate-mass models may be applicable to SNe~2005ap and 2008es,
which attained high peak luminosities and faded quickly as mentioned
above.  Again, an important consideration is that our models do not
include the possible delayed effects of diffusion when high optical
depths are important, as one might expect for massive and slow (and
therefore dense) CSM shells \citep{SM07}.  Including this may produce
a smoother light curve \citep{FA77}, especially in the sense that it
would dmpen and round-out the sharp initial peak in many of our
simulations.  Thus, it is possible that models such as B01 could be
dominated by diffusion and may not appear as SNe of Type IIn, but
confirming this conjecture requires additional work beyond the scope
of this paper.  Diffusion through an opaque shell, however, would not
alter the later phases of our light curves after maximum light.

There were some models that came close to matching the lightcurve
behavior of SNe~2006gy and 2006tf with both high luminosity and
relatively long durations.  These were models with SNe that had very
massive and extended shells of 10--25 $M_{\odot}$, although even these
seemed somewhat insufficient.  Figure~\ref{fig:compare} compares the
observed light curves of these two SNe (data from \citealt{S07,S08a})
to models A04, B02, and B04.  All of these have shell expansion speeds
of 200 km s$^{-1}$, close to the observed values of pre-shock material
for SNe~2006tf and 2006gy \citep{S07,S08a}.

\subsubsection{SN~2006tf} 

The light curve of SN~2006tf shows a slow and steady decline from peak
luminosity, the approximate rate of which is reproduced in all three
models shown.  In our simulations, this decline rate is mainly due to
the deceleration of the post-shock gas as the SN ejecta sacrifice
energy to radiation.  An interesting result is that this decline rate
during the main light curve peak for model B04 roughly matches the
radioactive decay rate of $^{56}$Co, even though there is no
luminosity from radioactivity included in these simulations --- in
other words, luminosity from shock-CSM interaction alone can in some
cases mimic the radioactive decay rate.  (One might also expect a
similar decline from a steeper density gradient in the CSM shell, or a
different velocity/density law in the SN ejecta.  In SN~2006gy, for
example, recent evidence points to a Hubble-like expansion law in the
CSM \citep{S09b}. However, none of the models sustain the high
luminosity for a long-enough time.  SN~2006tf shows relatively high
luminosity above 10$^{42}$ ergs s$^{-1}$ even at very late times {\it
  more than 1 yr after peak}, consistent with a continuation of the
same decay rate, but all three models in Figure~\ref{fig:compare} drop
long before that time.  In our simulations, this drop occurs when the
forward shock exits the outer boundary of the dense shell and
continues into the lower-density exterior wind shed by the star before
it ejected the CSM shell.  A similar drop in luminosity was observed
in the light curve of SN~1994W, and was also attributed to the shock
overruning the outer boundary of a CSM shell \citep{Cetal04}.  This
sharp drop is not usually seen in the light curves of SNe~IIn,
however, suggesting that most SNe~IIn have more extended CSM shells.

We explored the effect that changing the outer shell boundary has on
the light curves.  Models B02 and B04, both shown in
Figure~\ref{fig:compare}, are identical up to the point when the
radius of the outer shell boundary is reached in model B02.  At this
time, occurring around day 100, the luminosity in model B02 plummets
as the shock runs out of dense CSM to interact with.  In model B04,
however, we simply continued the same shell properties to a larger
radius by having the shell ejection occur with the same mass-loss rate
over a longer time interval ($\Delta t$=5 yr instead of 2 yr, both
ending 2 yr before core collapse).  Thus, the slow decline from peak
luminosity continued at roughly the same rate until day $\sim$260,
when its forward shock reached the outer shell boundary and the
luminosity finally plummeted.  This larger outer radius required a
much larger shell mass, increased from 10~$M_{\odot}$ in model B02 to
25~$M_{\odot}$ in model B04.  The general shape and luminosity of
model B04 is similar to the 25~$M_{\odot}$ shell model that Woosley et
al.\ (2007) suggested for SN~2006gy, although \citet{S08a} noted that
it also fit the early light curve of SN~2006tf well.  A shell mass of
25~$M_{\odot}$ is near the limit of what one might believe from a
non-terminal stellar outburst if giant eruptions of LBVs like
$\eta$~Carinae are representative \citep{S03,SF}.  Shell masses beyond
25~$M_{\odot}$ also begin to press the most basic limitations of even
a very massive star's mass budget at the end of its life (see,
e.g.,\citealt{SO6}).

However, even the extremely massive shell of 25~$M_{\odot}$ in model
B04 cannot sustain a high luminosity long enough to account for the +1
yr observations of SN~2006tf (Fig.\ \ref{fig:compare}) because it
drops too soon.  Instead, the late-time luminosity of SN~2006tf seems
to continue the same slow decline rate.  The corresponding CSM shell
mass that this would imply (roughly 50~$M_{\odot}$) is staggering if
the continued high luminosity were the result of simply extending the
same shell to larger radii.  One way to avoid such implausibly high
shell mass would be to lower the density of the envelope but increase
the total SN explosion energy above 10$^{51}$ ergs.  Higher explosion
energy leads to faster SN ejecta speed, so consequently, higher
instantaneous luminosity can be achieved with lower shell densities.
A larger explosion energy also relieves some of the strain on the
efficiency of converting kinetic energy into light, since the total
radiated energy of SN~2006tf is almost 10$^{51}$ ergs.

\subsubsection{SN~2006gy}

This SN presents additional challenges, since the total radiated
energy actually exceeded 10$^{51}$ ergs \citep{S07, S09b}, requiring a
more energetic SN explosion no matter what the CSM properties are.
None of our models were able to achieve the combination of the high
peak luminosity and long duration of SN~2006gy, although our most
energetic SN explosion was only 2$\times$10$^{51}$ ergs.  Following
the arguments above for SN~2006tf, then, one might expect that models
with a more energetic explosion could match the light curve of
SN~2006gy without having implausibly high CSM mass \citep{WBH07}.  For
example, the smooth light curve shape and slow rise to maximum are
traits that were seen in our simulations with bipolar CSM shells, so
one can imagine that a set of parameters similar to model D02 but with
higher explosion energy may account for the light curve of SN~2006gy.
This will be explored in a future paper.  Diffusion from opaque
shocked shell may also lead to a smooth light curve appropriate for
SN~2006gy \citep{SM07}, and we have not included these high
optical depth effects in our simulations.

\section{Conclusion}
\label{sec-conclusion}

In this paper we describe the influence of massive circumstellar
shells on core-collapse SN lightcurves, with the primary motivation of
trying to understand the power source of extremely luminous Type~IIn
events and their relation to the diverse population of SNe~IIn.  We
show how these circumstellar shells can indeed create extreme peaks in
the luminosity such as have been observed in Type~IIn supernovae like
SN2006gy and SN2006tf.  The luminosity of these SNe would require
extreme amounts of $^{56}$Ni if they are powered by radioactive decay,
but if interactions in the CSM provide the power instead, then the
shell masses and speeds that are required have reasonable precedent
from observed properties of spatially resolved shells around nearby
massive stars (see \citealt{SO6} and references therein).

Our investigation is by no means exhaustive.  Pre-SN circumstellar
shells may have a wide range of masses, expansion speeds, and radii,
whereas we have adopted simplified shell geometries for illustrative
examples.  Additionally, the underlying SNe ejecta may have wide
diversity in explosion energy, mass, and ejecta speed.  In this
preliminary investigation, our approach has been to vary each of these
parameters individually to illustrate their influence on the light
curve rather than attempting to accurately model any individual SN.
We have attempted to find general ways to distinguish between
different kinds of shells, using trends in the observed shapes of the
lightcurves, their characteristic emission temperature, and observed
shock speeds.  We find that observations of the evolution of the shock
speed is necessary to help break the degeneracy in the other free
parameters, while observations of the speed of the pre-shock CSM help
considerably as well (see e.g.,
\citealt{S02,S07,S08a,S09a,S09b,Tetal08}).  This can be used to
analyze the mass-loss history of massive stars in the last years prior
to the explosion, which can be a powerful tool for studying the final
stages of stellar evolution.  Ultimately, we wish to know the physical
origin of these SN-precursor events.

The key result is that we confirm the large masses of circumstellar
shells hypothesized to account for some recent luminous SNe~IIn
\citep{SM07,S07,S08a,S09b,WBH07}, as well as the high mass and
explosion energy of the underlying SNe.  One can also produce a very
high peak luminosity with lower mass if the shell is slow and the SN
ejecta are fast, but a lower mass shell cannot yield both a high peak
luminosity and a long duration of $\ga$100 days seen in some luminous
SNe~IIn. In fact, we suspect that even larger shell masses or larger
explosion energies are needed to account for the observed light curves
of the most luminous SNe~IIn. Thus, more detailed attempts to model
individual objects will be the focus of a second paper in this series.

\subsection{Future developements}

Further research is required for quantitative analyses of observed
SN~IIn lightcurves and to extract reliable absolute values of shell
masses and SN explosion energies.  This must include adding the
luminosity contribution from the underlying SN photosphere (powered by
diffusion or radioactive decay) in cases where the CSM interaction
luminosity is not extremely high compared to the ejecta photosphere,
as well as using an improved treatment of post-shock cooling and
radiative transfer at high optical depths in order to more accurately
model the emergent radiation from the post-shock shells in these
simulations.  As noted by \citet{SM07} and \citet{S08a, S09b}, it is
likely that the CSM will be highly opaque, especially at the earliest
phases, so the effects of radiative diffusion should be taken into
account to properly model the emergent luminosity.  Finally, all our
simulations have adopted a Type II-P core-collapse SN density profile,
but other types of SNe with different density profiles need to be
investigated in a similar manner, since any type of SN can, in
principle, be a Type~IIn event if it runs into a dense H-rich
environment.

\section*{acknowledgments}

A.J.v.M.\ acknowledges support from NSF grant AST-0507581, from the
FWO, grant G.0277.08 and K.U.Leuven GOA/09/009.  N.S.\ was partially
supported by NASA through grants GO-10241 and GO-10475 from the Space
Telescope Science Institute, which is operated by AURA, Inc., under
NASA contract NAS5-26555, and through Spitzer grants 1264318 and 30348
administered by JPL.  We thank R.H.D.\ Townsend for the use of his
radiative cooling module.  A.J.v.M.\ thanks G.\ Garc{\'i}a-Segura, N.\
Langer and R.\ Kotak for helpful discussion.
We also thank our anonymous referee for his insightful comments, 
which helped us to improve our paper.

\appendix

\section{Luminosity}
\label{sec-lum_table}
This appendix contains a  sample of our luminosity tables. The full tables can be found online.

\begin{table*}
   \label{tab:lum}
    \begin{centering}
      \caption{luminosity values for simulation O01}
      \begin{tabular}{p{0.2\linewidth}l}
         \hline
         \noalign{\smallskip} time [sec] &  log10(L) [erg/s]  \\
         \noalign{\smallskip}
         \hline
         \noalign{\smallskip}
  1.0000000E+05  & 4.0854548E+01  \\
  2.0000000E+05  & 4.0681275E+01  \\
  3.0000000E+05  & 4.0674991E+01  \\
  4.0000000E+05  & 4.0714879E+01  \\
  5.0000000E+05  & 4.0557764E+01  \\
  6.0000000E+05  & 4.0541291E+01  \\
  7.0000000E+05  & 4.0630674E+01  \\
  8.0000000E+05  & 4.0439978E+01  \\
  9.0000000E+05  & 4.0538254E+01  \\
  1.0000000E+06  & 4.0346281E+01 \\
          \hline
         \noalign{\smallskip}
     \end{tabular}
  \end{centering}
\end{table*}

\section{Shock velocity}
\label{sec-vel_table}
This appendix contains a sample of our shock velocity tables. The full tables can be found online.

\begin{table*}
   \label{tab:vel}
    \begin{centering}
      \caption{Shock velocity values for simulation O01}
      \begin{tabular}{p{0.2\linewidth}ll}
         \hline
         \noalign{\smallskip} time [sec] &  V(volume averaged) [cm/s]  &   V(mass averaged) [cm/s]\\
         \noalign{\smallskip}
         \hline
         \noalign{\smallskip}
  2.0000000E+05  & 6.3296810E+08  & 7.0761081E+08  \\
  3.0000000E+05  & 6.1563164E+08  & 6.8375272E+08  \\
  4.0000000E+05  & 5.9150803E+08  & 6.5441661E+08  \\
  5.0000000E+05  & 5.7608001E+08  & 6.4455979E+08  \\
  6.0000000E+05  & 5.7573592E+08  & 6.3470001E+08  \\
  7.0000000E+05  & 5.6127737E+08  & 6.1689406E+08  \\
  8.0000000E+05  & 5.4696696E+08  & 6.0784766E+08  \\
  9.0000000E+05  & 5.3620586E+08  & 6.0075791E+08  \\
  1.0000000E+06  & 5.3759244E+08  & 5.9993685E+08  \\	
          \hline
         \noalign{\smallskip}
     \end{tabular}
  \end{centering}
\end{table*}

\label{lastpage}


\begin{thebibliography}{}

\bibitem[\protect\citeauthoryear{Aretxaga et al.}{1999}]{Aetal99}
Aretxaga, I. et al.\ 1999, MNRAS, 309, 343

\bibitem[\protect\citeauthoryear{Arnett}{1996}]{A96} Arnett, D.\ 1996,
  Supernovae and Nucleosynthesis: An Investigation of the History of
  Matter from the Big Bang to the Present, Princeton series in
  astrophysics, Princeton University Press, 41 William St.  Princeton,
  NJ 08540, United States

\bibitem[\protect\citeauthoryear{Chevalier}{1977}]{C77} Chevalier,
  R.A.\ 1977, ARA\&A, 15, 175

\bibitem[\protect\citeauthoryear{Chevalier}{2005}]{C05} Chevalier,
  R.A.\ 2005, ApJ, 619, 839

\bibitem[\protect\citeauthoryear{Chevalier \& Fransson}{1992}]{CF92}
  Chevalier, R.A., \& Fransson, C., 1992, ApJ, 395, 540 %% SN expansion model

\bibitem[\protect\citeauthoryear{Chevalier \& Fransson}{2008}]{CF08}
  Chevalier, R.A., \& Fransson, C.\ 2008, ApJL, 683, 135 %% shcock break-out radiation

\bibitem[\protect\citeauthoryear{Chevalier \& Oishi}{2003}]{CO03}
  Chevalier, R.A., \& Oishi, J.\ 2003, ApJL, 593, 23

\bibitem[\protect\citeauthoryear{Chugai}{2001}]{C01}
  Chugai, N.N.\ 2001, MNRAS, 326, 1448

\bibitem[\protect\citeauthoryear{Chugai \& Danziger}{1994}]{CD94}
  Chugai, N.N., \& Danziger, I.J.\ 1994, MNRAS, 268, 173

\bibitem[\protect\citeauthoryear{Chugai et al.}{2004}]{Cetal04} Chugai,
  N.N., et al.\ 2004, MNRAS, 352, 1213

\bibitem[\protect\citeauthoryear{Clark}{1996}]{c96} Clark, D.A.\ 1996,
  ApJ, 457,291

\bibitem[\protect\citeauthoryear{Davidson \& Humphreys}{1997}]{dh97} Davidson, K., \& Humphreys,
  R.M.\ 1997, ARAA, 35, 1

\bibitem[\protect\citeauthoryear{Dessart et al.}{2008}]{D08} Dessart, L. et al.
\ 2008, ApJ, 675, 644

\bibitem[\protect\citeauthoryear{Dwarkadas \& Owocki}{2002}]{DO02}
  Dwarkadas, V.V., \& Owocki, S.P.\ 2002, ApJ, 581, 1337

\bibitem[\protect\citeauthoryear{Dyson \& Williams}{1997}]{DW} Dyson, J.E. \& Williams, D.A. \ 1997, 
The Physics of the Interstellar Medium, Institute of Physics publishing, Dirac House, Temple Back, 
Bristol BS1 6BE, UK

\bibitem[\protect\citeauthoryear{Falk \& Arnett}{1977}]{FA77} Falk, S., \&
  Arnett, D.W.\ 1977, ApJS, 33, 515

\bibitem[\protect\citeauthoryear{Filippenko}{1997}]{F97} Filippenko,
  A.V.\ 1997, ARA\&A, 35, 309

\bibitem[\protect\citeauthoryear{Fransson}{2002}]{F02} Fransson, C., et
  al.\ 2002, ApJ, 572, 350

\bibitem[\protect\citeauthoryear{Fox et al.}{2009}]{F09} Fox, O., et
  al.\ 2009, ApJ, 691, 650

\bibitem[\protect\citeauthoryear{Gal-Yam \& Leonard}{2009}]{GL09}
  Gal-Yam, A., \& Leonard, D.C.\ 2009, Nature, 458,865

\bibitem[\protect\citeauthoryear{Germany}{2000}]{G00} Germany, L.M.,
  et al.\ 2000, ApJ, 533, 320

\bibitem[\protect\citeauthoryear{Gezari et al.}{2009}]{getal2009}
  Gezari, S., et al.\ 2009, ApJ, 690, 1313

\bibitem[\protect\citeauthoryear{Hamuy}{2003}]{H03} Hamuy, M.\ 2003,
  ApJ, 582, 905

\bibitem[\protect\citeauthoryear{Hillier et al.}{2001}]{H01} Hillier, D.J., 
Davidson, K, Ishibashi, K, \& Gull, T.\ 2001, ApJ, 553, 837

\bibitem[\protect\citeauthoryear{Kasen \& Bildsten}{2009}]{KB09}
  Kasen, D., \& Bildsten, L.\ 2009, in press (arXiv:0911.0680)

\bibitem[\protect\citeauthoryear{Kasen \& Woosley}{2009}]{KW09} Kasen,
  D., \& Woosley, S.E.\ 2009, ApJ, 703, 2205

\bibitem[\protect\citeauthoryear{Leonard et al.}{2002}]{L02} Leonard,
  D.C., et al. \ 2002, PASP, 114, 35L

\bibitem[\protect\citeauthoryear{MacDonald \& Bailey}{1981}]{MB81}
  MacDonald, J., \& Bailey, M.E.,\ 1981, MNRAS, 197, 995

\bibitem[\protect\citeauthoryear{Matzner \& McKee}{1999}]{MM99}
  Matzner, C.D. \& McKee, C.F.\ 1999, ApJ, 510, 379

\bibitem[\protect\citeauthoryear{Miller et al.}{2009}]{metal2009}
  Miller, A.A., et al.\ 2009, ApJ, 690, 1303

\bibitem[\protect\citeauthoryear{Montes et al.}{1998}]{M98} Montes,
  M.J., Van Dyk, S.D., Weiler, K.W., Sramek, R.A., \& Panagia, N.\
  1998, ApJ, 506, 874

\bibitem[\protect\citeauthoryear{Ofek et al.}{2007}]{O07} Ofek, E.O.,
  et al.\ 2007, ApJL, 659, 13

\bibitem[\protect\citeauthoryear{Owocki}{2005}]{o05} Owocki, S.P.\
  2005, ASPC, 332, 169, proceedings of: The Fate of the Most Massive Stars, ASP Conference Series, Vol. 332, Proceedings of the conference held 23-28 May, 2004 in Grand Teton National Park, Wyoming. Edited by R. Humphreys and K. Stanek. San Francisco: Astronomical Society of the Pacific, 2005 p.171

\bibitem[\protect\citeauthoryear{Owocki et al.}{2004}]{OGS04} Owocki, S.P., 
Gayley, K.G. \& Shaviv, N.J.\ 2004, ApJ, 616, 525 %% clumping

\bibitem[\protect\citeauthoryear{Owocki \& Cohen}{2007}]{OC07} 
Owocki, S.P., \& Cohen, D.H.\ 2007, ApJ, 648, 565%% clumping

\bibitem[\protect\citeauthoryear{Prieto}{2007}]{P07} Prieto, J., et
  al.\ 2007, preprint (arXiv:0706L.4088)

\bibitem[\protect\citeauthoryear{Quimby et al.}{2007}]{qetal07}
  Quimby, R.M., et al.\ 2007, ApJL, 668, 99

\bibitem[\protect\citeauthoryear{Salamanca}{2002}]{S02} Salamanca, I., 
Terlevich, R.J. \& Tenorio-Tagle, G.\ 2002, MNRAS, 330, 844

\bibitem[\protect\citeauthoryear{Schlegel}{1990}]{S90} Schlegel, E.M.\
  1990, MNRAS, 244, 269

\bibitem[\protect\citeauthoryear{Smith}{2005}]{S05} Smith, N.\ 2005,
  MNRAS, 357, 1330

\bibitem[\protect\citeauthoryear{Smith}{2006}]{S06} Smith, N.\ 2006,
  ApJ,644,1151

\bibitem[\protect\citeauthoryear{Smith \& Ferland}{2007}]{SF} Smith,
  N., \& Ferland, G.\ 2007, ApJ, 655, 911

\bibitem[\protect\citeauthoryear{Smith \& Hartigan}{2006}]{SH06} Smith,
  N., \& Hartigan, P.\ 2006, ApJ, 638, 1045


\bibitem[\protect\citeauthoryear{Smith \& McCray}{2007}]{SM07} Smith,
  N., \& McCray, R.\ 2007, ApJL, 671, 17

\bibitem[\protect\citeauthoryear{Smith \& Owocki}{2006}]{SO6} Smith,
  N., \& Owocki, S.P.\ 2006, ApJL, 645, 45

\bibitem[\protect\citeauthoryear{Smith \& Townsend}{2007}]{ST07}
  Smith, N., \& Townsend, R.H.D.\ 2007, ApJ, 666, 967

\bibitem[\protect\citeauthoryear{Smith et al.}{2003}]{S03} Smith, N.,
  et al.\ 2003, AJ, 125, 1458

\bibitem[\protect\citeauthoryear{Smith et al.}{2007}]{S07} Smith, N.,
  et al.\ 2007, ApJ, 666, 1116 % 06gy

\bibitem[\protect\citeauthoryear{Smith et al.}{2008a}]{S08a} 
Smith, N., et al.\ 2008a, ApJ, 686, 467 % 06tf

\bibitem[\protect\citeauthoryear{Smith et al.}{2008b}]{S08b} Smith, N.,
  Foley, R.J., \& Filippenko, A.V.\ 2008b, ApJ, 680, 568 % 06jc

\bibitem[\protect\citeauthoryear{Smith et al.}{2008c}]{S08c} Smith,
  N., et al.\ 2008c, ApJ, 686, 485 % 06gy late time echo

\bibitem[\protect\citeauthoryear{Smith et al.}{2009}]{S09a} Smith, N.,
  et al.\ 2009, ApJ, 695, 1334

\bibitem[\protect\citeauthoryear{Smith et al.}{2010}]{S09b} Smith, N.,
  et al.\ 2010, ApJ, 709, 856

\bibitem[\protect\citeauthoryear{Stone \& Norman}{1992}]{SN92} Stone,
  J.M. \& Norman, M.L.\ 1992, ApJ, 80, 753

\bibitem[\protect\citeauthoryear{Townsend}{2009}]{t09} Townsend, R.H.D
  \ 2009, ApJS, 181, 391

\bibitem[\protect\citeauthoryear{Trundle et al.}{2008}]{Tetal08}
  Trundle, C., Kotak, R., Vink, J.S. \& Meikle,  W.P.S. \ 2008, A\&AL, 483, 47

\bibitem[\protect\citeauthoryear{Turatto et al.}{1993}]{T93}
  Turatto, M., et al.\ 1993, MNRAS, 262, 128

\bibitem[\protect\citeauthoryear{Van Dyk et al.}{1993}]{V93} Van Dyk,
  S.D., Weiler, K.W., Sramek, R.A., \& Panagia, N.\ 1993, ApJ, 419,
  L69

\bibitem[\protect\citeauthoryear{van Veelen et al.}{2009}]{vVetal09}
  van Veelen, B. et al.  A\&A, 503, 495

\bibitem[\protect\citeauthoryear{Whalen et al.}{2008}]{wetal08}
  Whalen, D., van Veelen, B., O'shea, B.W. \& Norman, M.L. \ 2008,
  ApJ, 682, 49

\bibitem[\protect\citeauthoryear{Williams et al.}{2002}]{W02}
  Williams, C.L., Panagia, N., Van Dyk, S.D., Lacey, C.K., Weiler,
  K.W., \& Sramel, R.A.\ 2002, ApJ, 581, 396

\bibitem[\protect\citeauthoryear{Woosley}{2009}]{W09} Woosley, S.E.\
  2009, in press (arXiv:0911.0698)

\bibitem[\protect\citeauthoryear{Woosley, Blinnikov \&
    Heger}{2007}]{WBH07} Woosley, S.E., Blinnikov, S. \& Heger, A.\
  2007, Nature, 450, 390

\bibitem[\protect\citeauthoryear{Young}{2004}]{Y04}
  Young, T.R.\ 2004, ApJ, 617, 1233


\end{thebibliography}
\end{document}